\documentclass[3p]{elsarticle}
\usepackage{epstopdf}
\usepackage{subfigure,subcaption,graphicx,float}
\usepackage{amsmath, amsfonts, amssymb}
\usepackage{amsthm}
\usepackage{yhmath}
\usepackage{epsf}
\usepackage{amsfonts}
\usepackage{stmaryrd}
\usepackage{amssymb}
\usepackage{leftidx}
\usepackage{xcolor}
\usepackage{mathtools}
\usepackage{placeins}
\usepackage{booktabs}
\usepackage{enumitem}
\usepackage{caption}
\usepackage{tabu}
\usepackage{multirow}
\usepackage{stackengine}
\usepackage[utf8]{inputenc}
\usepackage[english]{babel}
\usepackage{bm}
\usepackage{array}
\usepackage{multirow}

\def\bfu{{\bf u}}

\def\bfx{{\bf x}}

\def\bfE{{\bf E}}

\def\bfI{{\bf I}}

\def\bfN{{\bf N}}

\def\bfX{{\bf X}}

\def\eps{\varepsilon}

\def\e0{\varepsilon_0}

\def\s0{\sigma_0}

\def\sts{\sigma_{\texttt{ts}}}
\def\scs{\sigma_{\texttt{cs}}}
\def\shs{\sigma_{\texttt{hs}}}
\def\de{\delta^\varepsilon}
\def\ce{c_{\texttt{e}}}

\DeclareMathAlphabet{\mathsfit}{T1}{\sfdefault}{\mddefault}{\sldefault}
\SetMathAlphabet{\mathsfit}{bold}{T1}{\sfdefault}{\bfdefault}{\sldefault}

\theoremstyle{plain}
\newtheorem{theorem}{Theorem}

\newtheorem{remark}[theorem]{Remark} %SYMBOL DEFINITIONS

\long\def\symbolfootnote[#1]#2{\begingroup%
\def\thefootnote{\fnsymbol{footnote}}\footnote[#1]{#2}\endgroup}

\begin{document}
\begin{frontmatter}

\title{A comparison of phase field models of brittle fracture incorporating strength: I---Mixed-mode loading \vspace{0.1cm}}

\vspace{-0.1cm}

\author{Umar Khayaz}
\ead{ukhayaz3@gatech.edu}

\author{Aarosh Dahal}
\ead{adahal8@gatech.edu}

\author{Aditya Kumar\corref{cor1}}
\ead{aditya.kumar@ce.gatech.edu}

\address{School of Civil and Environmental Engineering, Georgia Institute of Technology, Atlanta, GA 30332, USA \vspace{0.05cm}}

\cortext[cor1]{Corresponding author}

\begin{abstract}

\vspace{-0.1cm}

The classical variational phase-field model for brittle fracture has proven highly effective in predicting the growth of large pre-existing cracks. Although challenges remain in accurately predicting crack evolution under anti-plane shear and compressive loading, the variational formulation based on Griffith's energy competition is widely accepted. However, the precise modeling of crack nucleation continues to be a significant challenge. Crack nucleation under uniform stress depends on the material's strength surface, an independent material property whose description is fundamentally incompatible with the energy-based Griffith propagation criterion. 
To address this, three main phase-field approaches have emerged, each attempting to reconcile material strength and toughness. The first, known as the classical variational approach, preserves the variational structure but fails to accurately incorporate the strength surface. In contrast, the other two approaches—the complete nucleation and hybrid cohesive zone models—sacrifice variational consistency. Among these, only the complete nucleation approach precisely accounts for the strength surface.  All three approaches, especially the second one, deviate from the sharp variational theory of brittle fracture, raising concerns about their reliability in predicting the growth of cracks under non-mode-I loading. 
This paper evaluates precisely this issue. It is the first in a series of studies comparing the three approaches, systematically investigating crack growth under mode II, mode III, and mixed-mode loadings. The results are validated against experimental and analytical benchmarks and compared with each other. The results confirm that the complete nucleation approach effectively predicts crack growth across all investigated problems, and its predictions agree well with those from other two approaches for tension-dominated cases.
Additionally, the findings highlight that an inaccurate accounting of the strength surface in the classical variational approach can potentially influence crack path predictions.
Lastly, they reveal that the modification of the crack driving force in the phase-field evolution equation to incorporate the strength surface in the hybrid cohesive zone approach results in crack propagation at an incorrect value of fracture toughness.

% , the incorporation of strength into the classical phase-field models necessarily affects the propagation as well. This is not well understood. Broadly, three different approaches for incorporating strength have been proposed. Two of them are based on ideas from the literature cohesive zone modeling and gradient damage modeling. The third idea takes a fresh approach and deviates from the variational or quasi-variational nature of the other two approaches. It has been argued to be superior for modeling nucleation of fracture. The goal of this paper is to compare the three approaches. A secondary goal is to provide further evidence for the ability of the non-variational third approach to describe accurately the crack propagation under mixed-mode loadings.
% A recent work has discussed the issues with crack nucleation in the classical variational models and cohesive zone models. In this paper, we show that the classical variational models, such as the star-convex model, fail even for crack propagation under mode II or III loading.

\endkeyword

\end{abstract}

\end{frontmatter}

\section{Introduction}

Francfort and Marigo's reformulation \cite{Francfort98} of Griffith's idea of an energetic competition between bulk strain energy and surface fracture energy \cite{Griffith21} governing both when large crack nucleates and how it propagates has led to a powerful theoretical and computational approach known as the phase-field method \cite{Bourdin00}. This method has been explored for extension beyond its original brittle fracture domain to address quasi-brittle fracture, ductile fracture, fatigue, and fracture influenced by other physical stimuli \cite{wu2017, borden2016ductile, dolbow2021ductile, kalina2023fatigue, svolos2020thermal, bourdin2019hydraulic}. However, several fundamental problems remain unresolved in the basic context of brittle fracture under quasi-static loading. The chief among them is the issue of modeling fracture nucleation in the bulk in regions distant from large cracks.
% , that is, the problem of fracture nucleation away from a large crack.

Macroscopic fracture nucleation in the bulk of the material under uniform stresses is governed by the strength of the material \cite{KBFLP20}. The strength is an independent macroscopic material property of brittle materials, albeit a stochastic one, due to the complexity of defect distribution at smaller scales and the challenges in maintaining uniform stress during measurement. Strength is determined experimentally by applying uniform stress to a sufficiently large\footnote{Large is defined with respect to the size of the intrinsic heterogeneity in the material as well as the intrinsic fracture length scales.}, homogeneous specimen. Only under uniform stresses does the material fail when a scalar function of the stress tensor reaches a critical threshold, called the strength surface \cite{KBFLP20}. In two or three dimensions, the strength surface is written as:
\begin{equation*}
\mathcal{F}(\boldsymbol{\sigma})=0,
\end{equation*}
where $\mathcal{F}$ represents a scalar function and $\boldsymbol{\sigma}$ is the stress tensor. While this condition suffices under uniform stress, predicting fracture nucleation under non-uniform stress requires knowledge of both the strength and the fracture toughness. 
Importantly, the strength surface, $\mathcal{F}(\boldsymbol{\sigma})=0$, cannot generally be reformulated in terms of energy or strain, although specific loading scenarios, such as uniaxial tension, permit equivalent energy or strain thresholds. For instance, critical strain or energy thresholds are often used in tensile testing of elastomers due to ease of measurement \cite{suo2017fractocohesive}, but these thresholds fail under hydrostatic loading conditions \cite{KFLP18, KLP21, KKLP24}.
This fundamental incompatibility between the stress-based criterion for nucleation and the energy-based Griffith criterion for propagation hinders the development of a unified model capable of addressing arbitrary loading conditions and geometries. Bridging the gap between strength and toughness remains a significant challenge in brittle fracture mechanics.

Historically, two solutions have been proposed to reconcile the strength and toughness criteria. 
The first approach retains Griffith’s variational formulation but replaces Griffith’s surface energy with cohesive surface energy \cite{Bourdin08,conti2016phase}.
The cohesive approach was motivated by a fictitious physical representation of crack faces bridged by linear or nonlinear springs \cite{barenblatt1959, dugdale1960}. The springs lose stiffness after reaching a macroscopic stress threshold, which can be related to the strength. Based on this physical representation, the energy dissipation in the springs, as represented by the cohesive surface energies, is taken to evolve with the crack opening displacement and traction at crack faces. The cohesive models also incorporate a threshold for crack opening displacement besides the stress threshold.
% , it does not affect predictions for brittle fracture.
While incorporating the strength surface, $\mathcal{F}(\boldsymbol{\sigma})=0$, is not typically attempted, the cohesive models are commonly extended to account for shear response in addition to the tensile response \cite{needleman1990analysis, xu1993void, park2009unified}. The second historical approach employs continuum damage theory, introducing an internal length scale to model nonlocal effects and indirectly embedding tensile strength into the formulation \cite{bazant1988nonlocal,lorentz1999variational, lorentz2017cohesive}. With these gradient damage approaches, various models to account for asymmetry in the critical stress threshold in tension and compression have been developed \cite{mazars1986split, mazars1989split, comi2001split, badel2007split}; however, general formulations that can account for the entire strength surface have largely not been studied.

These two ideas have been adopted into phase-field models to embed strength, primarily focusing on uniaxial tensile strength. Amor et al. \cite{AmorMarigoMaurini2009}, Pham and Marigo \cite{Pham2010a, Pham2010b} and Pham et al. \cite{pham2011gradient} proposed a reinterpretation of the phase-field models as gradient damage models where phase-field regularization length was assumed to be fixed and equal to the internal fracture length scale. The internal length scale was then related to the tensile strength under suitable scaling conditions. These models termed the classical variational models for fracture nucleation, were validated under mode I am loading conditions comprehensively by Tanne et al. \cite{Tanne18} and others. Following a cohesive approach, Lorentz \cite{lorentz2017cohesive} and Wu \cite{wu2017} proposed a class of models in which an internal length scale is introduced into the variational model through a modification of the elastic energy degradation function.  These models can essentially capture the various tensile traction-separation cohesive laws.

Extending these phase-field models to accurately capture the full-strength surface, and not just one point on the surface, remains underdeveloped. De Lorenzis and Maurini \cite{LorenzisMaurini2022nucleation} and Vicentini et al. \cite{LorenzisMaurini2024nucleation} have argued that carrying out suitable decompositions of the strain energy density function may help in accounting for more than one point on the strength surface. Currently, the representation of the strength surface with these decompositions is inadequate \cite{LDFL2025} due to the inherent difficulty in reformulating the strength surface as an energetic threshold.  For cohesive phase-field models, Lorentz \cite{lorentz2017cohesive} and Wu \cite{wu2017, wu2020vonMises} have proposed a hybrid formulation in which the energetic driving force for crack evolution is assumed to arise from an energy function separate from the strain energy function. They relate this new energy function with a damage threshold function. Zolesi and Maurini \cite{ZolesiMaurini24} have recently studied related cohesive models for nucleation under multiaxial stresses. 

Kumar et al. \cite{KFLP18, KBFLP20} have questioned the ability of classical variational and cohesive formulations to account for the entire strength surface and proposed a new approach.
Unlike the other methods, their approach argues for a formulation that is not variationally consistent and in which the strength surface is directly incorporated as a driving force in the evolution equation for phase-field instead of attempting to rewrite it in an energetic form. Quite remarkably, this approach has been shown to be successful in describing nucleation of fracture across a wide range of materials, loading conditions, and geometries, outperforming previous formulations in many scenarios \cite{KRLP22, KLDLP23, KKLP24, LK24}. The numerical evidence also includes several studies on crack propagation \cite{KBFLP20, KLP20, KKLP24}. However, questions have been raised about whether departing from the variational structure disrupts Griffith-type crack propagation under mixed-mode loading conditions.

Moreover, a comprehensive comparison of the three phase-field approaches that incorporate strength surfaces has yet to be conducted. This study seeks to fill this gap by systematically evaluating their ability to predict both nucleation and propagation from pre-existing cracks under various multiaxial and mixed-mode loading scenarios. Special attention is given to cases involving a prominent compressive strain field, where spurious crack formation has been a persistent issue, and incorporation of tension-compression asymmetry has been a topic of focus \cite{AmorMarigoMaurini2009, miehe2010, wick2022, LK24}. Additionally, we seek to examine whether incorporating the strength surface in the variational classical and cohesive models has an effect on crack propagation under general loading conditions. 
% A secondary goal of this study is to illustrate that the formulation of Kumar et al. \cite{KBFLP20,KKLP24} is able to describe crack propagation accurately under general loading conditions.

This paper focuses largely on cases where pre-existing cracks are present, with limited exploration of nucleation in other scenarios. Future parts of this study will address crack nucleation in bulk in different geometries and loading conditions. Furthermore, comparisons are restricted to linear elasticity, as classical variational and cohesive approaches incorporating strength are less developed for finite elasticity. Moreover, the evaluation is limited to the most recent models within each approach: the \emph{star-convex} model for classical variational phase-field \cite{LorenzisMaurini2024nucleation}, a linear softening cohesive phase-field model with modified von Mises equivalent stress \cite{wu2020vonMises, lorentz2017cohesive}, and the model of Kumar et al. in its latest form \cite{KKLP24}.  These models are described in Section 2, followed by numerical comparisons and discussions in Section 3, and final comments in Section 4.

% Following the work of Kumar et al., new variationally consistent formulations have been proposed in recent years attempting to describe the multiaxial strength surface. So-called hybrid formulations, that are variationally inconsistent, have also been introduced that modify the crack driving energy to describe different strength envelopes. However, these new formulations, as well as the formulation of Kumar et al., have raised the question of whether the modifications to the governing equations to account for strength leave the crack propagation undisturbed. This is precisely the question this paper seeks to answer.

% We solely focus here on the study of crack nucleation of propagation from a large or small pre-existing crack.
% We make no evaluation of the capability of any of these models to describe nucleation of fracture in any other case. We refer the reader to other work for that. The study of crack propagation presented in this work, however, is one of the most comprehensive studies encompassing analytical and experimental validation for various multiaxial and mixed-mode fracture tests. It is also the first comparison between the three different classes of approaches used in the literature to account for strength in phase-field methods.

\section{Phase-field approaches to brittle fracture incorporating strength}

\subsection{Initial configuration, kinematics, and material inputs}

Consider a structure made of an isotropic linear elastic brittle material occupying an open bounded domain $\mathrm{\Omega}\subset \mathbb{R}^3$, with boundary $\partial\mathrm{\Omega}$ in its undeformed and stress-free configuration at time $t=0$. At a later time $t \in (0, T]$, due to an externally applied displacement $\overline{\bfu}(\bfX, t)$ on a part $\partial\mathrm{\Omega}_\mathcal{D}$ of the boundary and a traction $\overline{\textbf{t}}(\bfX,t)$ on the complementary part $\partial\mathrm{\Omega}_\mathcal{N}=\partial\mathrm{\Omega}\setminus \partial\mathrm{\Omega}_\mathcal{D}$, the structure experiences a deformation field $\bfu(\bfX,t)$. We write the infinitesimal strain tensor as
\begin{equation*}
\bfE(\bfu)=\dfrac{1}{2}(\nabla\bfu+\nabla\bfu^T).
\end{equation*}
Non-interpenetration constraint implies that ${\rm det}(\bfI+\nabla\bfu)>0$.  In response to the externally applied mechanical stimuli, cracks can also nucleate and propagate in the structure. Those are described in a regularized way by a phase field
\begin{equation*}
v=v(\bfX,t)
\end{equation*}
taking values in $[0,1]$. Precisely, $v=1$ identifies regions of the sound material, whereas $0\le v<1$ identifies regions of the material that have been fractured.

Based on decades of experimental observations, the mechanical behavior of a brittle material is assumed to be fully characterized by three intrinsic properties of the material: (\emph{i}) elasticity, (\emph{ii}) critical energy release rate, and (\emph{iii}) strength. The elasticity for an isotropic linear elastic material is characterized by the stored-energy function
\begin{equation}
	W(\bfE(\bfu)) =\mu \, {\rm tr}\,\bfE^2+\dfrac{\lambda}{2}({\rm tr}\,\bfE)^2,\label{W-mu}
\end{equation}
where $\mu>0$ and $\lambda>-2/3\mu$ are the Lam\'e constants. Recall the basic relations $\mu=E/(2(1+\nu))$ and $\lambda=E\nu/((1+\nu)(1-2\nu))$, where $E$ is the Young's modulus and $\nu$ is the Poisson's ratio. The stress-strain relation is given by
\begin{equation*}
	\boldsymbol{\sigma}(\bfX,t)=\dfrac{E}{1+\nu}\bfE+\dfrac{E\,\nu}{(1+\nu)(1-2\nu)}({\rm tr}\,\bfE)\bfI.%\label{S-E-E}
\end{equation*}
The critical energy release rate (or fracture toughness), denoted as $G_c$, controls the nucleation from a large pre-existing crack through the Griffith criterion
\begin{equation}
-\dfrac{\partial{\mathcal{W}}}{\partial \Gamma}=G_c \label{Griffith},
\end{equation}
which describes a competition between bulk elastic energy and surface fracture energy. The strength of the material controls the nucleation in large specimens of homogeneous brittle materials subjected to a uniform state of stress $\boldsymbol{\sigma}$. In greater than one dimension, the set of critical stresses defines a surface in the stress space
\begin{equation}
\mathcal{F}(\boldsymbol{\sigma})=0,\label{SSurf-0}
\end{equation}
where $\boldsymbol{\sigma}$ stands for the Cauchy stress tensor. This surface is called the strength surface \cite{KBFLP20}. A crack will form in an indeterminate location in the homogeneous specimen subjected to arbitrary uniform loading once the stress hits the strength surface in any direction. Under non-uniform stress states, experimental observations have shown that the violation of the strength surface is not a sufficient condition for crack nucleation. In that case, an \emph{interpolation} of the strength criterion (\ref{SSurf-0}) and toughness criterion (\ref{Griffith}) defines the fracture nucleation. 

\begin{figure}[h!]
	\centering
	\includegraphics[width=3.in]{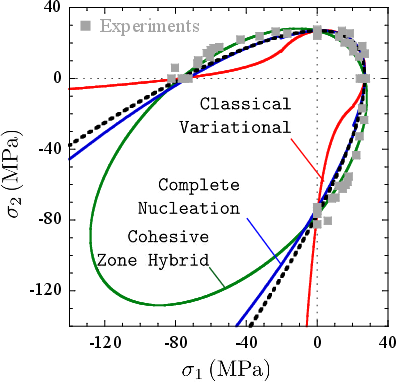}
	\caption{Comparison of the strength surfaces generated by the \texttt{Complete Nucleation} model (\ref{strength-surface-ce}) for $\varepsilon$=0.1 mm, \texttt{Classical Variational} model with star-convex energy decomposition (\ref{strength-surface-star-convex}), and \texttt{CZM-Hybrid} model (with modified von Mises stress) with the experimental data for graphite as reported by Sato et al. and the Drucker-Prager strength surface (black dashed line). All strength surface are plotted in principal stress space ($\sigma_1$, $\sigma_2$) with $\sigma_3=0$.}\label{Fig1}
\end{figure}

A popular choice for the strength surface of the material that we will invoke in this work is the Drucker-Prager strength surface
\begin{equation}
	\mathcal{F}(\boldsymbol{\sigma})=\sqrt{J_2}+\gamma_1 I_1+\gamma_0=0\qquad {\rm with}\qquad \left\{\hspace{-0.1cm}\begin{array}{l}\gamma_0=-\dfrac{2\sigma_{\texttt{cs}}\sigma_{\texttt{ts}}}
		{\sqrt{3}\left(\sigma_{\texttt{cs}}+\sigma_{\texttt{ts}}\right)}\vspace{0.2cm}\\
		\gamma_1=\dfrac{\sigma_{\texttt{cs}}-\sigma_{\texttt{ts}}}
		{\sqrt{3}\left(\sigma_{\texttt{cs}}+\sigma_{\texttt{ts}}\right)}\end{array}\right. ,\label{DP}
\end{equation}
where
\begin{equation}
	I_1={\rm tr}\,\boldsymbol{\sigma}\qquad {\rm and}\qquad  J_2=\dfrac{1}{2}{\rm tr}\,\boldsymbol{\sigma}^2_{D}\qquad {\rm with}\quad \boldsymbol{\sigma}_{D}=\boldsymbol{\sigma}-\dfrac{1}{3}({\rm tr}\,\boldsymbol{\sigma})\bfI\label{T-invariants}
\end{equation}
stand for two of the standard invariants of the stress tensor $\boldsymbol{\sigma}$, while the constants $\sigma_{\texttt{ts}}>0$ and $\sigma_{\texttt{cs}}>0$ denote the uniaxial tensile and compressive strengths of the material. This two-material-parameter Drucker-Prager strength surface (1952) is arguably the simplest model that has proven capable of describing reasonably well the strength of many nominally brittle materials. It describes well the experimentally measured strength data for isotropic graphite \cite{sato1987graphite} as shown in Fig.~\ref{Fig1}. We will primarily use graphite as the material of choice for numerical comparisons in this paper. 

Accepting this description of kinematics and material behavior, the key question is how to write down the balance equations that can capture the \emph{interpolation} of the strength criterion (\ref{SSurf-0}) and toughness criterion (\ref{Griffith}), and hence completely define the fracture nucleation. Here, different approaches and models have been developed that are discussed in the sequel.

{
\begin{remark}{\rm Some materials that are brittle in tension may show permanent deformation in compression before failure due to granular flow, pore collapse, crushing, and fragmentation, among other reasons. In that case, the surface (\ref{SSurf-0}) for $I_1<0$ may represent the onset of the permanent deformation. Localization and increasing permanent deformation may follow, and crack formation may occur when a bounding surface in the stress-deformation space is reached.
}
\end{remark}
}

\subsection{The classical variational approach}\label{Sec:variational}

Variational phase-field models were developed by Bourdin et al. \cite{Bourdin00} as an approximation of the variational theory of brittle fracture to study crack growth following Griffith's postulate. They are of the form
\begin{eqnarray}
&(\bfu^\eps_k,v^\eps_k)=\underset{\begin{subarray}{c}
  \bfu=\overline{\bfu}(t_k)\,{\rm on}\,\partial\Omega_\mathcal{D} \\[1mm]
  0 \leq v\leq v_{k-1}\leq 1
  \end{subarray}}{\arg\min}\,\mathcal{E}^{\eps}(\bfu,v):=\displaystyle
  \int_{\Omega} g(v) W(\bfE(\bfu))\,{\rm d}\bfx+\dfrac{G_c}{4 c_s}\int_{\Omega}\left(\dfrac{s(v)}{\eps}+\eps\nabla v\cdot\nabla v\right)\,{\rm d}\bfx,\label{BFM00}
\end{eqnarray}
where $\varepsilon$ is a regularization length,  $g$ and $s$ are continuous strictly monotonic functions, called degradation and surface regularization functions respectively, such that $g$(0) = 0, $g$(1) = 1, $s$(0) = 1, $s$(1) = 0, and $c_s = \int_0^1 \sqrt{s(z)}{\rm d}z$  is a normalization parameter. The generally preferred choice for the degradation and surface regularization functions are $g(v)=v^2$ and $s(v)=1-v$, collectively known as the \texttt{AT$_1$} model. A residual stiffness $\eta$ is often added to the degradation function for stability, $g(v)=(v^2+\eta)$.  Numerical simulations in crack growth with this model showed crack growth under compressive strain/stress fields. It was argued that the root cause of this behavior is the lack of impenetrability constraint or unilateral contact in the formulation \cite{Bourdin00, AmorMarigoMaurini2009, chambolle2018interpenetration}. As a remedy, Amor et al. \cite{AmorMarigoMaurini2009} proposed a remedy to split the free energy as
\begin{equation}
\mathcal{W}=v^2 W^{+}(\bfE(\bfu))+ W^{-}(\bfE(\bfu)),
\end{equation}
where a ``tensile" part of the energy,  $W^{+}(\bfE(\bfu)$, is only degraded with the phase field, and the ``compressive" part, $W^{-}(\bfE(\bfu))$, is not. This split is not unique and often fails to resolve issues with impenetrability and contact, as further discussed in Remark \ref{Remark-contact}. Many energy splits have been proposed in the literature. Incorporating the split and traction boundary conditions, the governing  system of coupled partial differential equations (PDEs) can be obtained from the variational principle above for
the displacement field $\bfu_k(\bfX)=\bfu(\bfX,t_k)$ and phase field $v_k(\bfX)=v(\bfX,t_k)$ at any material point $\bfX\in\overline{\mathrm{\Omega}}$ and discrete time $t_k\in\{0=t_0,t_1,...,t_m,$ $t_{m+1},...,$ $t_M=T\}$ are
\begin{equation}
\left\{\begin{array}{ll}
 {\rm Div}\left[v_{k}^2\dfrac{\partial W^{+}}{\partial \bfE}(\bfE(\bfu_{k})) + \dfrac{\partial W^{-}}{\partial \bfE}(\bfE(\bfu_{k})) \right]={\bf0},\quad \bfX\in\mathrm{\Omega},\\[10pt]
\bfu_{k}=\overline{\bfu}(\bfX,t_{k}),\quad \bfX\in\partial  \mathrm{\Omega}_{\mathcal{D}},\\[10pt]
 \left[v_{k}^2\dfrac{\partial W^{+}}{\partial \bfE}(\bfE(\bfu_{k})) + \dfrac{\partial W^{-}}{\partial \bfE}(\bfE(\bfu_{k}))\right]\bfN=\overline{\textbf{t}}(\bfX,t_{k}),\quad \bfX\in\partial \mathrm{\Omega}_{\mathcal{N}}\end{array}\right. \label{BVP-u-variational}
\end{equation}
and
\begin{equation}
\left\{\begin{array}{l}
\dfrac{3}{4} \varepsilon \,  \,  G_c \triangle v_{k}=2 v_{k} W^{+}(\bfE(\bfu_{k}))- \dfrac{3}{8}  \dfrac{ \, G_c}{\varepsilon},
 \mbox{if } v_{k}(\bfX)< v_{k-1}(\bfX),\quad \bfX\in\mathrm{\Omega} \\[10pt]
\dfrac{3}{4} \varepsilon \,  \,  G_c \triangle v_{k}\geq2 v_{k} W^{+}(\bfE(\bfu_{k}))- \dfrac{3}{8} \dfrac{ \, G_c}{\varepsilon},
 \mbox{if } v_{k}(\bfX)=1\; \mbox{ or }\; v_{k}(\bfX)= v_{k-1}(\bfX)>0,\quad \bfX\in\mathrm{\Omega} \\[10pt]
v_{k}(\bfX)=0,\quad \mbox{ if } v_{k-1}(\bfX)=0,\quad \bfX\in\mathrm{\Omega}
\\[10pt]
\nabla v_{k}\cdot\bfN=0,\quad \bfX\in \partial\mathrm{\Omega}
   \end{array}\right. \label{BVP-v-variational}
\end{equation}
with $\bfu(\bfX,0)\equiv\textbf{0}$ and $v(\bfX,0)\equiv1$, where $\nabla\bfu_k(\bfX)=\nabla\bfu(\bfX,t_k)$, $\nabla v_k(\bfX)=\nabla v(\bfX,t_k)$, $\triangle v_k(\bfX)= \triangle v(\bfX,$ $t_k)$.

The system of equations (\ref{BVP-u-variational})-(\ref{BVP-v-variational}) is only suitable to study crack growth when a large crack is already present in the specimen under study. Motivated by the similarity of this formulation to the more developed gradient damage formulations at that time, Amor et al. \cite{AmorMarigoMaurini2009} followed by Pham and Marigo \cite{Pham2010a, Pham2010b} and Pham et al. \cite{pham2011gradient} proposed to view the regularization length $\varepsilon$ as a material length scale. They showed through a one-dimensional analysis that for a fixed value of length $\varepsilon$, an initially uniform phase field $v(t)$ will lose stability at a critical stress for sufficiently large geometry. The critical stress will be the uniaxial tensile strength of the material obtained through the following equation in terms of the fixed value of $\varepsilon$:
\begin{equation}
\sts^2=\dfrac{3 G_c E}{8 \eps}.
\end{equation}
The fixed value of $\eps$ is termed the characteristic fracture length scale, $l_{\rm ch}$, which was regarded as an intrinsic material property.
While they noted that this formulation with their proposed energy split would lead to a nearly symmetric response in tension and compression, attempts to capture an asymmetric response or, in general, capture the entire strength surface were not pursued. 

% A material whose strength surface is described by the Rankine criterion, max($\sigma_1, \sigma_2, \sigma_3,$)=$\sts$ can, however, be described well by the approach above.

Following the work of Kumar et al. \cite{KBFLP20} that showed the critical importance of incorporating experimentally consistent strength surfaces such as the Drucker-Prager type surface into phase-field models, several attempts were made to incorporate Drucker-Prager type strength surfaces into the variational models \cite{LorenzisMaurini2022nucleation, paneda2022general}. The idea was to make use of alternative energy splits that could better describe the strength surfaces. Recall that the energy splits were originally not developed for this purpose but rather as a remedy for enforcing the unilateral contact condition. As shown recently by Vicentini et al. \cite{LorenzisMaurini2024nucleation}, the approach of splitting energy to capture the Drucker-Prager surface turns out to be ineffectual because an arbitrary split of energy can result in models that do not have crack-like residual stresses. Also, this approach can not remedy the issue of crack nucleation in incompressible materials under hydrostatic loading raised in \cite{KFLP18, KLP21, KKLP24}.

Still, Vicentini et al. \cite{LorenzisMaurini2024nucleation} have proposed another ad-hoc energy split that has been argued to be superior to the previous splits and possesses the desired tension-compression asymmetry in strength. They split the energy into tensile and compressive parts as follows:
\begin{equation}
\left\{\begin{array}{l}
W^{+}(\bfE(\bfu)) =\mu \, \left({\rm tr}\,\bfE^2 - \dfrac{1}{3} ({\rm tr}\,\bfE)^2 \right)+\dfrac{\kappa}{2}\left(({\rm tr}\,\bfE^{+})^2 - \gamma^{\star} ({\rm tr}\,\bfE^{-})^2 \right)\\[10pt]
W^{-}(\bfE(\bfu)) = (1+\gamma^{\star}) \dfrac{\kappa}{2} ({\rm tr}\,\bfE^{-})^2 
   \end{array}\right. \label{star-convex-split}
\end{equation}
where ${\rm tr}\,\bfE^{+}=({\rm tr}\,\bfE+|{\rm tr}\,\bfE|)/2$ and ${\rm tr}\,\bfE^{-}=({\rm tr}\,\bfE-|{\rm tr}\,\bfE|)/2$, $\kappa= \lambda + (2/3) \mu$ is the bulk modulus, and $\gamma^{\star}>-1$ is a constant that is related to the ratio of compressive strength and tensile strength through the prescription
\begin{equation}
\gamma^{\star}=\frac{1}{1 - 2\nu} \left( 2(1 + \nu) - \frac{3 \sts^2}{\scs^2} \right). \label{gamma-SC}
\end{equation}
The \texttt{AT}$_1$ variational model with this split was dubbed the \emph{star-convex} model.  For $\gamma^{\star}=0$, the star-convex model reduces to the volumetric-deviatoric model of Amor et al. \cite{AmorMarigoMaurini2009}. The strength surface generated for star-convex  model can be written as \cite{LDFL2025}
\begin{equation}
\mathcal{F}^{\rm SC}(\boldsymbol{\sigma})=\left\{\begin{array}{l}
\dfrac{\sqrt{J_2}}{\mu}+\dfrac{I_1^2}{9 \kappa}- \dfrac{3 G_c}{8 \eps}=0, \qquad I_1>0 \\[10pt]
\dfrac{\sqrt{J_2}}{\mu}- \gamma^{\star} \dfrac{I_1^2}{9 \kappa}- \dfrac{3 G_c}{8 \eps}=0, \qquad I_1<0
   \end{array}\right. \label{strength-surface-star-convex}
\end{equation}
The strength surface is shown in comparison to the strength data for graphite in Fig.~\ref{Fig1}. It is plain to see that there is a significant disagreement, especially in the fourth quadrant, first noted in \cite{LDFL2025}. Moreover, as the star-convex model is a generalization of the volumetric-deviatoric model---which has been shown to encounter challenges in accurately predicting crack evolution under mixed mode loading \cite{zhang2022assessment,LK24}---it is uncertain whether the star-convex model might face similar issues. It is this aspect that we primarily evaluate in Section \ref{Sec: Benchmark}.

{
\begin{remark}{\rm Accounting for the material non-interpenetration constraint and the crack-like residual stress behavior in compression in regularized phase-field models remains an unresolved challenge. While these two requirements are often considered synonymous, they are fundamentally distinct. The non-interpenetration constraint is primarily a kinematic condition, meaning it can be satisfied without necessarily ensuring accurate contact behavior.
Such is the case for the volumetric-deviatoric split \cite{AmorMarigoMaurini2009}. It has been shown that it approximates well in $\Gamma-$convergence sense the non-interpenetration constraint \cite{chambolle2018interpenetration}. However, 
this split does not show the correct contact behavior in compression as shown by Steinke and Kaliske \cite{steinke2019} and others. This is due to its incompressible fluid-like behavior in compression, which is not representative of a crack.  Another widely adopted energy decomposition in literature is the spectral split \cite{miehe2010}. This split was designed purely to drive fracture in regions of positive strains. While it can transfer stresses correctly in compression, it shows unphysical residual stresses in shear \cite{steinke2019, LorenzisMaurini2024nucleation, LK24}. Strobl and Seelig \cite{stroblseelig2015} and Steinke and Kaliske \cite{steinke2019} have proposed a so-called directional split as a remedy; see Fan et al. \cite{wick2022} for a review of different variants of this split. In this method, stresses and strain are decomposed into their normal and shear components with respect to a local crack coordinate system \cite{steinke2019}. Based on this decomposition, energy is computed for \emph{linear} elastic materials.  However, this decomposition can only be performed in a fully formed phase-field crack at some distance behind the crack tip \cite{stroblseelig2015}. To predict crack evolution, the directional split requires \emph{a priori} knowledge of the crack orientation, which nullifies the main advantage of the phase-field method in predicting arbitrary crack evolution without using any additional criteria. In general, there does not exist a rational way to decompose an arbitrary strain energy density function itself into ``tensile" and ``compressive" parts in two or three space dimensions.
} \label{Remark-contact}
\end{remark}
}

{
\begin{remark}{\rm The inequalities in (\ref{BVP-v-variational}) describe the two constraints on the phase field, namely that its value remains between 0 and 1 and that it decreases monotonically in time. The second constraint is due to the fact that fracture is an irreversible process in most scenarios. The constraints are accounted for using the penalty method in this work \cite{KFLP18, KLDLP23}.
}
\end{remark}
}

\subsection{The phase-field approach of Kumar et al. (2018, 2020)}\label{Sec: Kumar}

The phase-field approach of Kumar et al. \cite{KFLP18, KBFLP20} was motivated by the observation that in the classical variational approach, strength is considered subordinate to elasticity and toughness, and this prevents an accurate accounting of the strength surface (\ref{SSurf-0}) in the phase-field model. They proposed two generalizations of the variational approach:
\begin{enumerate}
\item to consider the Euler-Lagrange equations of the phase-field regularization---and not the variational principle (\ref{BFM00}) itself---as the primal model,
\item to account for the strength surface directly through the addition of a stress-based driving force in the Euler–Lagrange equation governing the evolution of the phase field rather than indirectly through an energy split.
\end{enumerate}
In this complete nucleation approach, the displacement field $\bfu_k(\bfX)=\bfu(\bfX,t_k)$ and phase field $v_k(\bfX)=v(\bfX,t_k)$ at any material point $\bfX\in\overline{\mathrm{\Omega}}$ and discrete time $t_k\in\{0=t_0,t_1,...,t_m,$ $t_{m+1},...,$ $t_M=T\}$ are determined by the system of coupled partial differential equations (PDEs)
\begin{equation}
\left\{\begin{array}{ll}
 {\rm Div}\left[v_{k}^2\dfrac{\partial W}{\partial \bfE}(\bfE(\bfu_{k}))\right]={\bf0},\quad \bfX\in\mathrm{\Omega},\\[10pt]
\bfu_{k}=\overline{\bfu}(\bfX,t_{k}),\quad \bfX\in\partial  \mathrm{\Omega}_{\mathcal{D}},\\[10pt]
 \left[v_{k}^2\dfrac{\partial W}{\partial \bfE}(\bfE(\bfu_{k}))\right]\bfN=\overline{\textbf{t}}(\bfX,t_{k}),\quad \bfX\in\partial \mathrm{\Omega}_{\mathcal{N}}\end{array}\right. \label{BVP-u-theory}
\end{equation}
and
\begin{equation}
\left\{\begin{array}{l}
\dfrac{3}{4} \varepsilon \, \de \,  G_c \triangle v_{k}=2 v_{k} W(\bfE(\bfu_{k}))+c_\texttt{e}(\bfX,t_{k})- \dfrac{3}{8}  \dfrac{\de \, G_c}{\varepsilon},
 \mbox{if } v_{k}(\bfX)< v_{k-1}(\bfX),\quad \bfX\in\mathrm{\Omega} \\[10pt]
\dfrac{3}{4} \varepsilon \, \de \,  G_c \triangle v_{k}\geq2 v_{k} W(\bfE(\bfu_{k}))+c_\texttt{e}(\bfX,t_{k})- \dfrac{3}{8} \dfrac{\de \, G_c}{\varepsilon},
 \mbox{if } v_{k}(\bfX)=1\; \mbox{ or }\; v_{k}(\bfX)= v_{k-1}(\bfX)>0,\quad \bfX\in\mathrm{\Omega} \\[10pt]
v_{k}(\bfX)=0,\quad \mbox{ if } v_{k-1}(\bfX)=0,\quad \bfX\in\mathrm{\Omega}
\\[10pt]
\nabla v_{k}\cdot\bfN=0,\quad \bfX\in \partial\mathrm{\Omega}
   \end{array}\right. \label{BVP-v-theory}
\end{equation}
where $c_\texttt{e}(\bfX,t)$ is the additional driving force containing information about material's strength and $\de$ is a non-negative coefficient that ensures the crack propagation behavior is consistent with the Griffith's postulate. For a suitable constitutive prescription of $c_\texttt{e}$ discussed below, the model becomes length-scale independent, and thus $\varepsilon$ acts as purely a regularization length scale.
The constitutive prescription for $\ce$ depends on the particular form of the strength surface. The basic recipe for the construction of $\ce$ for arbitrary strength surface was presented in \cite{KBFLP20}; see also \cite{senthilnathan2025} for an updated discussion and derivation of $\ce$ for Mohr-Coulomb strength surface. It is spelled out next for the case of Drucker-Prager strength surfaces (\ref{DP}). 

The driving force $\ce$ takes the same functional form as the Drucker-Prager strength surface (\ref{DP}) but with $\varepsilon$-dependent coefficients. Precisely, it is defined as
\begin{align}
c_{\texttt{e}}(\bfX,t)=\widehat{c}_{\texttt{e}}(I_1,J_2;\varepsilon)
=\beta_2^\varepsilon\sqrt{\mathcal{J}_2}+\beta_1^\varepsilon \mathcal{I}_1,\label{cehat-2020}
\end{align}
where
\begin{equation}
\left\{\begin{array}{l}
\beta^\varepsilon_1=\dfrac{1}{\shs}\delta^\varepsilon\dfrac{G_c}{8\varepsilon}-\dfrac{2\mathcal{W}_{\texttt{hs}}}{3\shs}\vspace{0.2cm}\\
\beta^\varepsilon_2=\dfrac{\sqrt{3}(3\shs-\sts)}{\shs\sts}\delta^\varepsilon\dfrac{G_c}{8\varepsilon}+
\dfrac{2\mathcal{W}_{\texttt{hs}}}{\sqrt{3}\shs}-\dfrac{2\sqrt{3}\mathcal{W}_{\texttt{ts}}}{\sts}\end{array}\right. , \label{betas}
\end{equation}
with
$$\mathcal{W}_{\texttt{ts}}= \dfrac{\sts^2}{2 E}, \quad \mathcal{W}_{\texttt{hs}}= \dfrac{\shs^2}{2 \kappa}, \quad \shs=\dfrac{2 \sts  \scs} { 3 (\scs - \sts)}, \quad \kappa= \dfrac{E}{3 (1 - 2 \nu)}.$$
$\mathcal{I}_1$ and $\mathcal{J}_2$ stand for the invariants (\ref{T-invariants}) of the degraded Cauchy stress
\begin{equation*}
\boldsymbol{\sigma}(\bfX,t)=v^2\dfrac{\partial W}{\partial \bfE}(\bfE(\bfu))
\end{equation*}
and, hence, read as
\begin{equation*}
\mathcal{I}_1=(3\lambda+2\mu) v^2 {\rm tr}\,\bfE(\bfu)\quad {\rm and}\quad \mathcal{J}_2=2\mu^2 v^4 {\rm tr}\,\bfE^2_D(\bfu)
\end{equation*}
with $\bfE_D(\bfu)=\bfE(\bfu)-1/3\left({\rm tr}\,\bfE(\bfu)\right)\bfI$ in terms of the displacement field $\bfu$ and phase field $v$. The formulas for the two constants $\beta^\varepsilon_1$ and $\beta^\varepsilon_2$ are obtained by fitting the strength surface exactly at two points---in this case, the uniaxial tensile strength and hydrostatic tensile strength.

Remarkably, with this constitutive choice, the system of equations (\ref{BVP-u-theory})-(\ref{BVP-v-theory}) show behavior consistent with the Griffith's equation (\ref{Griffith}); however, with a different effective critical energy release rate. The value of the effective critical energy release rate can be corrected to match the experimental value, $G_c$, through the parameter $\de$. The parameter $\de$ can be numerically calibrated for any boundary-value problem of choice for which the nucleation from a large pre-existing crack can be determined exactly according to Griffith’s equation (\ref{Griffith}). An approximate analytical formula for $\de$ was recently provided in Kamarei et al. \cite{KKLP24}
\begin{equation}
\delta^\varepsilon=\left(1+\dfrac{3}{8}\dfrac{h}{\varepsilon}\right)^{-2}\left(\dfrac{\sts+(1+2\sqrt{3})\,\shs}{(8+3\sqrt{3})\,\shs}\right)\dfrac{3  G_c}{16 \mathcal{W}_{\texttt{ts}} \varepsilon}+\left(1+\dfrac{3}{8}\dfrac{h}{\varepsilon}\right)^{-1}\dfrac{2}{5}.
\label{delta-eps-final-h}
\end{equation}
where $h$ is the mesh size. The maximum error obtained with this approximate formula is 5-10 $\%$ in terms of the effective value of critical energy release rate for $\eps$ in the range $[l_{\rm ch}/10, 2 l_{\rm ch}]$ that was tested in previous studies.  This formula is applicable for any brittle material, whether described by linear elastic or nonlinear elastic theory, for which the strength surface is well captured by the Drucker-Prager strength surface (\ref{DP}). Based on the numerical results presented in various studies \cite{KBFLP20, KRLP22, KLDLP23, KKLP24, LK24}, it is also observed to be independent of the boundary value problem under investigation. We test this observation further in Section \ref{Sec: Benchmark} with more studies in mixed-mode fracture. 

The strength surface generated by this phase field model is given by the equation:
\begin{equation}
\mathcal{F}^{\rm CN}(\boldsymbol{\sigma})=
\dfrac{\sqrt{J_2}}{\mu}+\dfrac{I_1^2}{9 \kappa} + \ce - \dfrac{3 \de G_c}{8 \eps}=0 \label{strength-surface-ce}
\end{equation}
obtained by setting $v=1$ in the right-hand side of the evolution equation for phase field (\ref{BVP-v-theory}). The numerical results \cite{KBFLP20, KLP20} have indicated that for sufficiently large structures, the initially uniform phase field loses stability when the condition above is satisfied for arbitrary multiaxial loadings. The homogenous 1D solution for this model has been presented by Meng et al. \cite{najafi2025}. The strength surface $\mathcal{F}^{\rm CN}$ reduces to the exact strength surface for $\eps \searrow 0$ \cite{KBFLP20, senthilnathan2025}. However, for large values of localization lengths $\varepsilon$,  the strength surface approximation with $\mathcal{F}^{\rm CN}$ can be poor in compressive regions ($I_1<0$). A remedy was proposed in \cite{KRLP22} to add the following correction term to $\ce$ to improve the strength surface representation 
\begin{align}
c_{\texttt{e}}(\bfX,t)=\widehat{c}_{\texttt{e}}(I_1,J_2;\varepsilon)
=\beta_2^\varepsilon\sqrt{\mathcal{J}_2}+\beta_1^\varepsilon \mathcal{I}_1-
\dfrac{1}{v^3}\left(1-\dfrac{\sqrt{\mathcal{I}_1^2}}{\mathcal{I}_1}\right)\left(\dfrac{\mathcal{J}_2 (1+\nu)}{E}+\dfrac{\mathcal{I}_1^2 (1-2 \nu)}{6 E}\right),\label{cehat-2022}
\end{align}
where $\beta_1^\varepsilon, \beta_2^\varepsilon$ are still given by equations (\ref{betas}) and $\de$ has the same prescription (\ref{delta-eps-final-h}) as well. The strength surface generated with this choice of $\ce$ is shown in Fig.~\ref{Fig1} in comparison with experimental data for graphite and the exact Drucker-Prager strength surface. See also \cite{LK24, liu2024dynamic} for a detailed account of the correction term. In practice, the irreversibility is also only enforced once the phase field satisfies $v<0.05$ \cite{KBFLP20}.

\begin{remark}{\rm The computational implementation of the equations (\ref{BVP-u-theory})--(\ref{BVP-v-theory}) differs from the implementation of the classical variational model (\ref{BVP-u-variational})--(\ref{BVP-v-variational}) only through the presence of the term $\ce$ on the right-hand side of the PDE for the phase-field (\ref{BVP-v-theory}). Furthermore, it has been noted recently by Larsen et al. \cite{larsen2024} that the equations (\ref{BVP-u-theory})--(\ref{BVP-v-theory})  can be recast as a variational theory. Specifically, they have shown that the solution pair $({\bfu}, v)$ for the PDEs correspond to the fields that minimize separately two different functionals. One is the deformation energy functional
\begin{equation}
 \mathcal{E}^\varepsilon_d(\mathbf{u}_k; v_k) :=
\int_\Omega \left( v_k^2 + \eta_\varepsilon \right) W\left( \mathbf{E}(\mathbf{u}_k) \right) \, {\rm d}\bfx
- \int_{\partial \Omega_N} \bar{\mathbf{t}}(\mathbf{X}, t_k) \cdot \mathbf{u}_k \, {\rm d}\bfx.   
\end{equation}
and the second is the fracture functional
\begin{equation}
\mathcal{E}^\varepsilon_f(v_k; \mathbf{u}_k) :=
\int_\Omega v_k^2  W\left(\mathbf{E}(\mathbf{u}_k)\right) \, {\rm d}\bfx
+ \int_\Omega \frac{v_k^3}{3} \, \ce (\mathbf{E}(\mathbf{u}_k)) \, {\rm d}\bfx
+ \frac{3 \de G_c}{8} \int_\Omega \left( \frac{1 - v_k}{\varepsilon} + \varepsilon \nabla v_k \cdot \nabla v_k \right) \, {\rm d}\bfx.
\end{equation}
This alternating minimization process is no different from the solution process for the classical variational models.
}
\end{remark}

{
\begin{remark}{\rm 
Spurious branching of crack into regions of compressive strain is avoided in the phase-field approach of Kumar et al. without an energy split as shown recently in \cite{LK24}. This is due to the natural asymmetry introduced by the embedding of the strength surface as an independent material property. The compressive strength of brittle materials is typically significantly larger than the tensile strength. This asymmetry in strength is sufficient in preventing compressive cracks; see also the discussion in Liu et al. \cite{liu2024dynamic} in the context of dynamic fracture. However, it is not only the asymmetry of tensile and compressive strength that matters; the shape of the strength surface is also important, as discussed later in Section \ref{inclinednotch} by way of an example.

 On the other hand, the solution of the equations (\ref{BVP-u-theory})--(\ref{BVP-v-theory}) is suspect to material interpenetration under compression. While an energy split can be used as a partial remedy, it is not a comprehensive solution, as discussed in Remark \ref{Remark-contact}. A split is not adopted for the comparisons presented in the next section.
}
\end{remark}
}

\begin{remark}{\rm The fracture theory (\ref{BVP-u-theory})--(\ref{BVP-v-theory}) is a macroscopic theory. As such, it accounts not for the explicit presence of the inherent microscopic defects in the material but only for their macroscopic manifestation---the strength---via the driving force $\ce$. Since $\ce$ is a manifestation of defects that are not part of the macroscopic thermodynamic system, we interpret it as an \emph{external} force. In a configurational (micro-force) force framework \cite{Gurtin99}, $\ce$ appears as an external configurational force---parallel to how a body force appears in the balance of Newtonian forces---and is thermodynamically consistent. The interested reader is referred to \cite{KRLP18} for a complete account.
}
\end{remark}

\subsection{The phase-field cohesive zone approach} \label{Sec:CZM}

Motivated by a desire to make the classical variational approach length-scale independent and consistent with cohesive crack zone models, Lorentz \cite{lorentz2011cohesive, lorentz2017cohesive} proposed to adopt the following choice for the degradation function $g(v)$:
\begin{equation}
g(v)= \dfrac{v^2}{v^2+ a_1 (1-v) P(v) } \label{Lorentz-degradation}
\end{equation}
for $a_1>0$ and a continuously differentiable, strictly
positive and bounded function $P(v)$, such that $P(1)=1$. Different choices of the function $P(v)$ leads to different softening behavior. A specific choice for the function $P(v)$ was suggested to be $P(v)=1+ a_2 (1-v)$, where $a_2$ is another constant. Wu \cite{wu2017} later suggested higher-order polynomials for the function $P(v)$ to capture more complex softening behavior. They also suggested a different surface regularization function $s(v)=1-v^2$. For this choice of $s(v)$ and a linear softening behavior, they obtained the following values for the constants $a_1$ and $a_2$ 
\begin{equation}
a_1= \dfrac{2 E G_c}{\sts^2} \dfrac{2}{\pi \varepsilon}  \quad {\rm and} \quad a_2=-0.5. \label{Wu-a1a2}
\end{equation}
Plugging the choices above into the variational principle (\ref{BFM00}) and obtaining the Euler-Lagrange equations lead to a $\eps$-independent model that can capture the Griffith behavior correctly as well as the nucleation under uniaxial tension. However, similar to the classical $\texttt{AT}_1$ model without an energy split, it ignores the rest of the strength surface. Lorentz \cite{lorentz2017cohesive} proposed a remedy to write the damage driving force, $\mathcal{Y} (\bfE, v)$, in the evolution equation for the phase field as
\begin{equation}
 \mathcal{Y} (\bfE, v) = g'(v) \, \Gamma(\bfE),  \label{drivingforce}
\end{equation}
in place of the variationally consistent choice, $\mathcal{Y} (\bfE, v) = g'(v) W(\bfE)$, for the classical model. The function $\Gamma(\bfE)$ is constructed based on a damage threshold function $f(\bfE)$. Wu \cite{wu2017} adopted the same idea but proposed to write the driving force in terms of the stress instead of strain
\begin{equation}
 \hat{\mathcal{Y}} (\boldsymbol{\sigma}, v) = g'(v) \,\Gamma(\boldsymbol{\sigma}).  \label{CZM-drivingforce}
\end{equation}

With these choices, the displacement field $\bfu_k(\bfX)=\bfu(\bfX,t_k)$ and phase field $v_k(\bfX)=v(\bfX,t_k)$ at any material point $\bfX\in\overline{\mathrm{\Omega}}$ and discrete time $t_k\in\{0=t_0,t_1,...,t_m,$ $t_{m+1},...,$ $t_M=T\}$ are determined by the system of coupled partial differential equations (PDEs)
\begin{equation}
\left\{\begin{array}{ll}
 {\rm Div}\left[g(v_{k}) \dfrac{\partial W}{\partial \bfE}(\bfE(\bfu_{k}))\right]={\bf0},\quad \bfX\in\mathrm{\Omega},\\[10pt]
\bfu_{k}=\overline{\bfu}(\bfX,t_{k}),\quad \bfX\in\partial  \mathrm{\Omega}_{\mathcal{D}},\\[10pt]
 \left[g(v_{k}) \dfrac{\partial W}{\partial \bfE}(\bfE(\bfu_{k}))\right]\bfN=\overline{\textbf{t}}(\bfX,t_{k}),\quad \bfX\in\partial \mathrm{\Omega}_{\mathcal{N}}\end{array}\right. \label{BVP-u-Wu}
\end{equation}
and
\begin{equation}
\left\{\begin{array}{l}
\dfrac{2}{\pi} \varepsilon \,  \,  G_c \triangle v_{k}=g'(v_{k}) \, \Gamma (\boldsymbol{\sigma})- \dfrac{G_c}{\pi}  \dfrac{2 v_k}{\varepsilon},
 \mbox{if } v_{k}(\bfX)< v_{k-1}(\bfX),\quad \bfX\in\mathrm{\Omega} \\[10pt]
\dfrac{2}{\pi} \varepsilon \,  \,  G_c \triangle v_{k} \geq g'(v_{k}) \, \Gamma (\boldsymbol{\sigma})- \dfrac{G_c}{\pi}  \dfrac{2 v_k}{\varepsilon} ,
 \mbox{if } v_{k}(\bfX)=1\; \mbox{ or }\; v_{k}(\bfX)= v_{k-1}(\bfX)>0,\quad \bfX\in\mathrm{\Omega} \\[10pt]
v_{k}(\bfX)=0,\quad \mbox{ if } v_{k-1}(\bfX)=0,\quad \bfX\in\mathrm{\Omega}
\\[10pt]
\nabla v_{k}\cdot\bfN=0,\quad \bfX\in \partial\mathrm{\Omega}
   \end{array}\right. \label{BVP-v-Wu}
\end{equation}
where $g'(v_{k})= \frac{a_1 v_k (2 + 2 a_2 (1 - v_k) - v_k)}{\left(v_k^2+ a_1 (1-v_k) +a_1 a_2 (1-v_k)^2 \right)^2}$ based on (\ref{Lorentz-degradation}).
This is often dubbed a hybrid formulation, as the stress field and the crack driving force are derived from different energy functions.
A specific choice for $\Gamma (\boldsymbol{\sigma})$ based on the damage criterion introduced by de Vree et al. \cite{devree1995}, called the \emph{modified von Mises} criterion is commonly adopted in the literature \cite{lorentz2017cohesive, wu2020vonMises}
\begin{equation}
\Gamma (\boldsymbol{\sigma})= \dfrac{\sigma_{\rm eq}^2}{2 E}, \quad {\rm where} \quad    \sigma_{\rm eq}= \left(\dfrac{ \rho_c-1}{2 \rho_c}\right) I_1 + \dfrac{1}{2 \rho_c} \sqrt{ (\rho_c-1)^2 I_1^2 + 12 \rho_c J_2} \quad {\rm with} \quad \rho_c=\dfrac{\scs}{\sts}. \label{von Mises criterion}
\end{equation}
The strength surface generated for this  model is given by 
\begin{equation}
\mathcal{F}^{\rm CZM}(\boldsymbol{\sigma})=
a_1\dfrac{\sigma_{\rm eq}^2}{E}- \dfrac{3 G_c}{8 \eps}=0 \label{strength-surface-CZM}
\end{equation}
The strength surface is shown in comparison to the strength data for graphite in Fig.~\ref{Fig1}. It shows fairly good agreement with the experimental data. However, it predicts unphysical phase-field evolution under hydrostatic compressive stress states. Another specific choice for $\Gamma (\boldsymbol{\sigma})$ proposed in \cite{wu2017, wu2020vonMises} is the \emph{Rankine} criterion
\begin{equation}
\Gamma (\boldsymbol{\sigma})= \dfrac{\sigma_{\rm eq}^2}{2 E}, \quad {\rm where} \quad    \sigma_{\rm eq}= \dfrac{\sigma_{\rm max} + |\sigma_{\rm max}|}{2} \quad {\rm with} \quad \sigma_{\rm max}={\rm max} (\sigma_1, \sigma_2, \sigma_3). \label{Rankine} 
\end{equation}
The impact of using the hybrid formulation on the large-crack propagation under mixed-mode loading has not been studied before. It is this aspect that we focus on in the next section. Note that no energy split is employed with the cohesive zone hybrid formulation. Also note that a similar hybrid formulation idea can also be employed with the classical \texttt{AT$_1$} model.

\section{Benchmark problems and comparisons with experiments}\label{Sec: Benchmark}

We now present simulations of several benchmark problems using the three phase-field models introduced in the last section and compare their predictions. The benchmarks include the analysis of large-crack growth in notched plates under mode I, mode II, and mode III loading. Additionally, crack growth is examined for notched plates under biaxial tensile and pure shear loading conditions.
We also simulate the mode I+II+III benchmark problem from Yosibash and Mittleman \cite{yosibash2016, yosibash2020}, which includes extensive experimental data for validating the crack path. Finally, we investigate the fracture behavior of double-edge notched specimens under combined mode I+II loading, and a plate with inclined notch under compressive loading. 

We use the label \texttt{Complete Nucleation} to refer to the class of models due to Kumar et al. \cite{KBFLP20, KKLP24} discussed in Section \ref{Sec: Kumar}. This label is chosen due to the ability of these models to incorporate arbitrary strength surface as an independent material property. The Drucker-Prager surface is adopted for all simulations in this work. The label \texttt{Classical Variational} or simply \texttt{Variational} is used to refer to the models discussed in Section \ref{Sec:variational}.  These models adopt a fixed value of regularization length and an decomposition of the strain energy function to approximately describe the strength surface. Specifically, the star-convex model (\ref{star-convex-split}) is adopted for comparisons due to its superiority to previous variational approaches based on recent work by Vicentini et al. \cite{LorenzisMaurini2024nucleation}. Finally, we use the label \texttt{Cohesive Zone Hybrid} or simply \texttt{CZM-Hybrid} to refer to the class of phase-field models due to Lorentz \cite{lorentz2017cohesive} and Wu \cite{wu2017} discussed in Section \ref{Sec:CZM}. These models employ an alternative degradation function which allow to recover various softening behavior. A linear softening behavior (\ref{Wu-a1a2}) is assumed for comparisons in this work. They also take the crack driving force to depend on a separate energy function, $\Gamma$, which is related with a damage threshold function (\ref{CZM-drivingforce}).  The choice based on modified von Mises equivalent stress criterion (\ref{von Mises criterion}) is primarily used for the comparisons. 

\subsection{Mode-I surfing}

\begin{figure}[h!]
	\centering
	\includegraphics[width=5.5in]{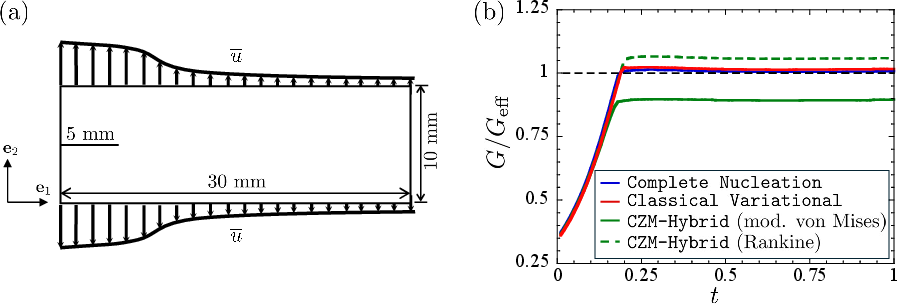}
	\caption{Propagation of a crack in mode I in a specimen subjected to ``surfing" boundary condition schematically depicted in (a). (b) Prediction of the normalized energy release rate as a function of time for three phase-field models, namely the \texttt{Complete Nucleation} model, the \texttt{Classical Variational} model, and the cohesive zone model with equivalent stress-based driving force (\texttt{CZM-Hybrid}). \texttt{CZM-Hybrid} is simulated with both modified von Mises and Rankine models.}\label{Fig2}
\end{figure}

First, we study the “surfing” boundary-value problem, introduced by Hossain et al. \cite{hossain2014}, to investigate the mode I crack propagation with  the three models. The basic idea behind the surfing problem
consists in subjecting a long strip of the material of interest with a pre-existing crack on its side to a suitably selected
boundary condition that makes the pre-existing crack propagate in a stable manner at a constant energy release rate. 
The schematic of the specimen geometry and boundary conditions is shown in Fig.~\ref{Fig2}. Specifically, we subject the top ($x_2=5$ mm) and bottom ($x_2=-5$ mm) boundaries of the strip to the displacement
\begin{align}
\overline{u}_2(\bfx,t)=\overline{u}(x_1-V t,x_2)   \label{Surfing-BC}
\end{align}
within the setting of plane stress, where the particular form of the function $\overline{u}$ is not critically important. Here, for definiteness, we make use of
\begin{align}
\overline{u}(x_1,x_2)=\dfrac{(1+\nu)\sqrt{G_c}}{\sqrt{2\pi E}}\left[x_1^2+x_2^2\right]^{1/4}\left[\dfrac{3-\nu}{1+\nu}-\cos \left(\tan^{-1}\left(\dfrac{x_2}{x_1}\right)\right)\right] \sin\left(\dfrac{1}{2}\tan^{-1}\left(\dfrac{x_2}{x_1}\right)\right), \label{Surfing-BC-u}
\end{align}

Both the \texttt{Classical Variational} and \texttt{Complete Nucleation} models have been extensively studied for this problem \cite{Tanne18, KBFLP20, KLP20, KKLP24}. However, to the authors' knowledge, the results for the \texttt{CZM-Hybrid} model have not been reported, providing the motivation for this study.
We adopt the material properties of graphite as reported by Sato et al. \cite{sato1987graphite} and Goggin and Reynolds \cite{goggin1967graphite}; see also Kumar et al. \cite{KBFLP20}. The Drucker-Prager strength surface provides an accurate description of the strength data for graphite, as illustrated in Fig.~\ref{Fig1}. Based on experimental data, Young's modulus $E$ and Poisson's ratio $\nu$ are taken as $9.8$ GPa and $0.13$, respectively. The critical energy release rate $G_c$ is set to $91$ N/m, with a tensile strength $\sts$ of $27$ MPa and a compressive strength $\scs$ of $77$ MPa. The regularization length scale $\varepsilon$ is chosen to be 0.25 mm for the \texttt{Complete Nucleation} and \texttt{CZM-Hybrid} models. For the \texttt{Classical Variational} model, $\varepsilon$ is set to the characteristic length in uniaxial tension $l_{\rm ch} \approx 0.46$ mm.   Simulations employ damaged notch boundary conditions, as described in \cite{Tanne18}. Fig.~\ref{Fig2}(b) reports the evolution of the energy release rate G in the strips for the three models, obtained by calculating the J-integral over the boundary of the strips. The energy release rate is normalized with respect to the effective fracture toughness. For the \texttt{Classical Variational} model (\ref{BVP-v-variational}), the effective fracture toughness is calculated as $G_{\rm eff}=G_c (1+\frac{3 h}{8 \eps})$ \cite{Tanne18}. In the case of the \texttt{CZM-Hybrid} model (\ref{BVP-v-Wu}), it is given by $G_{\rm eff}=G_c (1+\frac{h}{\pi \eps})$. For the \texttt{Complete Nucleation} model, the mesh size correction is incorporated directly into the definition of the parameter $\de$ (\ref{delta-eps-final-h}).

\begin{figure}[h!]
	\centering
	\includegraphics[width=4.5in]{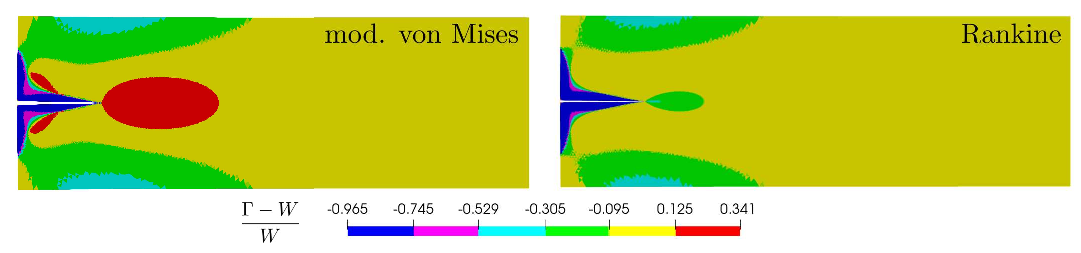}
	\caption{Contour plots of the normalized difference of the crack-driving energy function in the \texttt{CZM-Hybrid} model, $\Gamma$ with the strain energy function, $W$ from an elastic calculation without fracture. The normalized difference with the modified von Mises criterion is shown on left, and the one with the Rankine criterion is shown on the right.}\label{Fig3}
\end{figure}

The results presented in Fig.~\ref{Fig2}(b) indicate that both the \texttt{Classical Variational} and \texttt{Complete Nucleation} models successfully predict crack propagation with good accuracy. In contrast, the \texttt{CZM-Hybrid} model fails to provide a correct prediction of the fracture toughness. This inaccuracy arises from the model's incorporation of a different energy function $\Gamma(\boldsymbol{\sigma})$ to drive fracture. While this crack-driving function is equivalent to the strain energy under uniaxial tensile stress conditions, this equivalence does not hold in more general loading scenarios. The percentage difference between the crack-driving function, $\Gamma(\boldsymbol{\sigma})$ and the strain energy, $W(\bfE)$ is shown in Fig.~\ref{Fig3}. A substantial difference can be observed near the crack tip. Making use of the Rankine criterion in the \texttt{CZM-Hybrid} model improves the predictions of fracture toughness in Fig.~\ref{Fig2}(b). Nonetheless, the results exhibit a deviation of approximately 10\% in effective fracture toughness. This residual discrepancy can again be understood from the difference between $\Gamma(\boldsymbol{\sigma})$ and $W(\bfE)$ in Fig.~\ref{Fig3}.
This analysis highlights the limitations of the \texttt{CZM-Hybrid} approach of bringing the strength surface information into the phase-field model. A similar limitation can be inferred for many other models that use stress-based criteria for the crack driving force, such as those proposed by Miehe et al. \cite{miehe2015} and Bilgen and Weinberg \cite{weinberg2019stress}.

\subsection{Plate subjected to in-plane and anti-plane shear} \label{Mode2&3}

\begin{figure}[h!]
	\centering
	\includegraphics[width=6.5in]{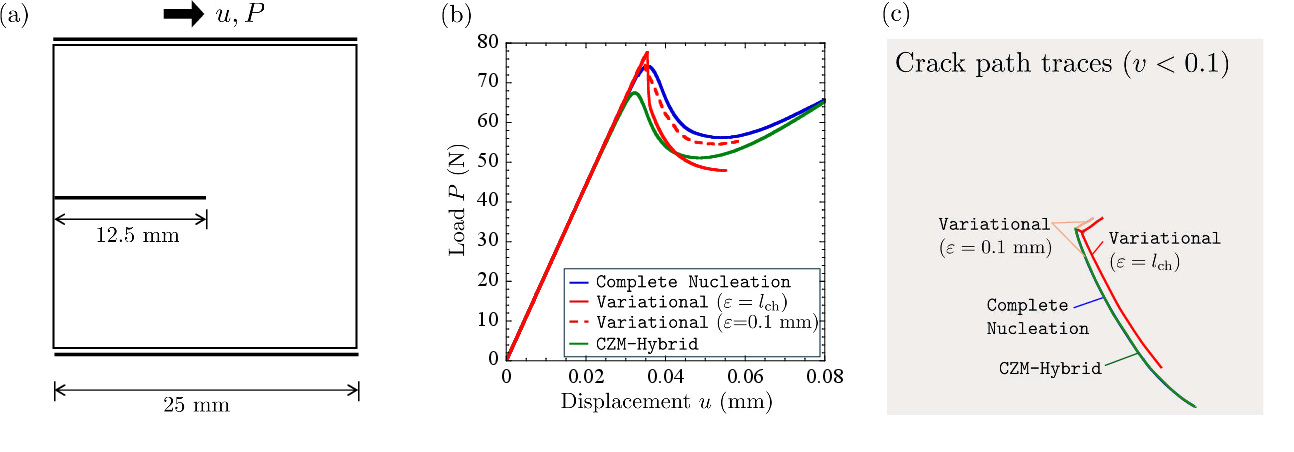}
	\caption{Comparison of the three phase-field models for a plate subjected to in-plane shear. (a) Schematic of the geometry and applied boundary conditions. (b) Plot of the predicted load $P$ as a function of the applied displacement $u$. (c) Plot tracing the crack path predicted by the three models. The \texttt{Classical Variational} results correspond to the star-convex model and show a compressive branch.}\label{Fig4}
\end{figure}

Simulating crack propagation under mode II and mode III ``surfing" loading is inherently challenging due to the tendency of cracks to deviate from the primary plane. Thus, we focus on studying plates subjected to in-plane and anti-plane shear conditions, comparing the force-displacement curves and crack paths predicted by the three phase-field models. The material properties of graphite, as defined earlier, are utilized for consistency.
A geometric crack is introduced under damaged notch boundary conditions. For the in-plane shear problem, a constant displacement is applied along the ${\bf e}_1$ direction at the top and bottom boundaries as shown in Fig.~\ref{Fig4}(a), while for the anti-plane shear problem, the displacement is applied in the ${\bf e}_3$ direction as illustrated in Fig.~\ref{Fig7}(a).

The problem of a plate under in-plane shear has been extensively studied in the literature with all three models. For the \texttt{Classical Variational} model without energy splitting, symmetric tensile and compressive cracks emanate from the existing crack. Various energy splitting methods, including the volumetric-deviatoric split, have been shown to eliminate the spurious compressive crack effectively. However, an unexpected result is observed with the star-convex split in our simulations---shown in Fig.~\ref{Fig4}(c). Initially, only the tensile crack nucleates and propagates, following a path nearly identical to that predicted by the \texttt{Complete Nucleation} model and the \texttt{CZM-Hybrid} model for an identical regularization length $\varepsilon=0.1$ mm. However, after a while, the \texttt{Classical Variational} model with the star-convex split predicts the nucleation of a short compressive crack originating from the existing crack. This behavior persists across different values of $\varepsilon$, which indirectly controls the material strength. For instance, results for $\varepsilon=l_{\rm ch}$ included in Fig.~\ref{Fig4}(c) also demonstrate the nucleation of a compressive crack. To rule out the possibility that the compressive crack is a numerical artifact caused by a distorted crack region, the calculations were repeated using a residual stiffness of $\eta=10^{-4}$, as suggested in \cite{AmorMarigoMaurini2009}. However, identical results were obtained. It is worth noting that the force-displacement response of the \texttt{Classical Variational} model with $\varepsilon=0.1$ mm closely matches the response of the \texttt{Complete Nucleation} model up until the nucleation of the compressive crack, as illustrated in Fig.~\ref{Fig4}(b). In contrast, the \texttt{CZM-Hybrid} model predicts premature crack nucleation due to an incorrect fracture toughness even though the crack path is only slightly affected.

\begin{figure}[h!]
	\centering
	\includegraphics[width=5.in]{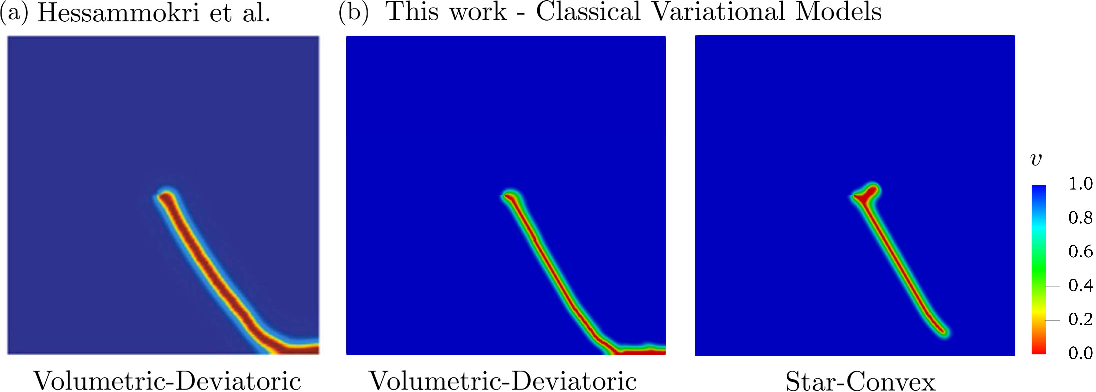}
	\caption{Contour plots of the phase field $v$ in the in-plane shear test using the \texttt{Classical Variational} model from (a) the work of Hessammokri et al. \cite{hesammokri2023} with volumetric-deviatoric energy split and (b) this work using the volumetric-deviatoric and star-convex splits.}\label{Fig5}
\end{figure}

The \texttt{Classical Variational} model with star-convex split is a generalization of the volumetric-deviatoric split, specifically designed to provide greater flexibility in setting compressive strength values. To validate our results, we conduct comparative simulations using both splitting approaches and compare them with previously published results \cite{hesammokri2023}, shown in Fig.~\ref{Fig5}. Geometry and material properties were adopted from \cite{hesammokri2023}. The domain was scaled down to 1 mm. Following material properties are taken: $E=210$ GPa, $\nu=0.3$, $G_c=2700$ N/m, $\sts=2$ GPa, and $\scs=6$ GPa.  Our findings reveal that the anomalous behavior manifests exclusively in the star-convex split.

This anomalous behavior can be explained through an examination of equation (\ref{star-convex-split}), which defines the star-convex split. When ${\rm tr}\,\bfE<0$, both splitting methods exhibit fluid-like characteristics, characterized by a large bulk-to-shear modulus ratio of $(1+\gamma^{\star}) \kappa / \eta$, where $\eta$ represents the residual stiffness and $\gamma^{\star}$ equals zero in the volumetric-deviatoric case. Although this fluid-like response helps in facilitating stress transmission under compression---see also Remark \ref{Remark-contact}---it simultaneously introduces an unintended consequence: the generation of artificial pressure at crack tips and crack corners. The numerical investigations indicate that this artificial pressure, amplified by the $\gamma^{\star}$ term, is the primary mechanism responsible for initiating compressive crack formation in the star-convex model.

\begin{figure}[h!]
	\centering
	\includegraphics[width=5.in]{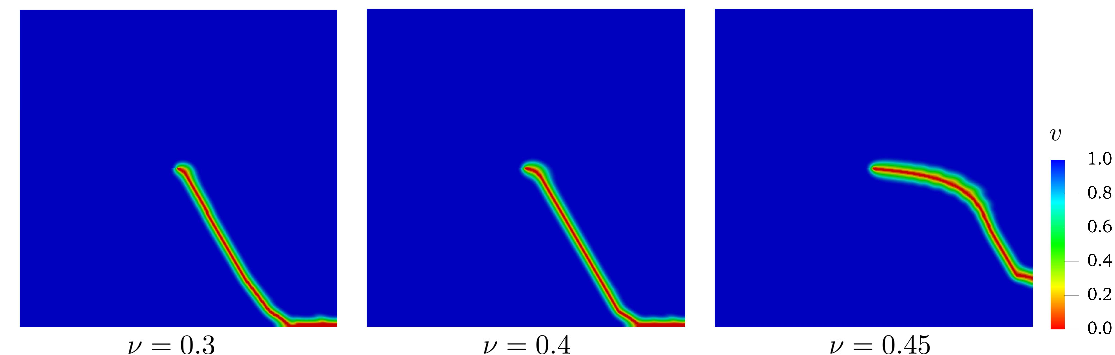}
	\caption{Contour plots of the phase field $v$ resulting from the \texttt{Classical Variational} model with volumetric-deviatoric split for three different values of the Poisson's ratio $\nu$.}\label{Fig6}
\end{figure}

Our comprehensive numerical studies, conducted across diverse parameter combinations, do not reveal the same pathological behavior with the volumetric-deviatoric model. However, given the structural similarities in the formulations, we cannot definitively rule out this possibility\footnote{In the Kalthoff dynamic fracture test, the nucleation of a secondary crack in compressive strain region with the volumetric-deviatoric split is well known \cite{zhang2022assessment,liu2024dynamic}}. Nevertheless, we found a distinct phenomenon where the predictions of the variational model with volumetric-deviatoric split deviate from the other two models. Specifically, as the material's Poisson's ratio $\nu$ increased, crack propagation exhibit a preference for shear initiation with the volumetric-deviatoric model as shown in Fig.~\ref{Fig6}. In contrast, both the \texttt{Complete Nucleation} and \texttt{CZM-Hybrid} models consistently produce tensile cracks across all tested values of $\nu$. We explore this distinction further in our analysis of the mode III problem next. The potential influence of Poisson's ratio on a tensile-to-shear crack transition in brittle materials would require further experimental investigation.

\begin{figure}[h!]
	\centering
	\includegraphics[width=6.5in]{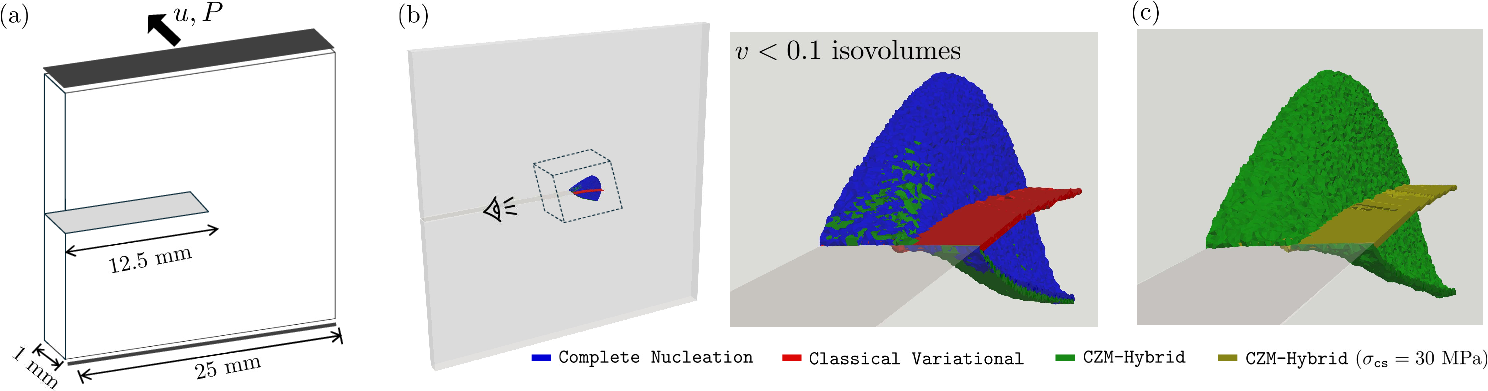}
	\caption{Comparison of the three phase-field models for a plate subjected to anti-plane shear. (a) Schematic of the 3D geometry and applied boundary conditions. (b)  Crack isovolumes computed for the values of the phase field in the range $0 \le v < 0.1$ as predicted by the three models for graphite with compressive strength $\sigma_\texttt{cs}=77$ MPa. (c) Crack isovolumes predicted by the \texttt{CZM-Hybrid} model for graphite with $\sigma_\texttt{cs}=77$ MPa and $\sigma_\texttt{cs}=30$ MPa.}\label{Fig7}
\end{figure}

The crack propagation under mode III loading is less understood than the mode II loading. Although various crack propagation criteria have been proposed in the literature, experimental observations indicate that cracks tend to deviate from their original plane to minimize or eliminate both mode II and mode III stress components. This out-of-plane rotation frequently results in crack front segmentation. Given that the underlying mechanics of crack segmentation are not yet fully comprehended, this investigation focuses on thin plates where such segmentation is unlikely to occur.
The analysis again employs graphite material properties and implements the three phase-field models with a regularization length $\varepsilon=0.1$ mm. 
Fig.~\ref{Fig7}(b) presents the results through overlapping isovolume where the phase-field parameter $v$ is less than 0.1.
The \texttt{Complete Nucleation} and \texttt{CZM-Hybrid} models yield nearly identical tensile crack paths characterized by out-of-plane rotation. In contrast, the \texttt{Classical Variational} model predicts a shear crack path that initially propagates along the original crack plane. The behavior of the three models remains qualitatively unchanged for larger plate thicknesses. This disparity between the variational model and other models appears analogous to the shear cracks observed at high Poisson ratios in Fig.~\ref{Fig6} and can likely be attributed to the artificially low shear strengths inherent in \texttt{Classical Variational} models.

To validate the hypothesis regarding the influence of realistic shear strength as a barrier against shear crack formation, additional simulations are conducted using the \texttt{Complete Nucleation} and \texttt{CZM-Hybrid} models with a reduced compressive strength of 30 MPa. This modification uniformly decreases strength values across shear and adjacent stress states. Under these conditions, both models produce shear crack paths comparable to those predicted by the \texttt{Classical Variational} model with $\scs=77$ MPa. Fig.~\ref{Fig7}(c) demonstrates these results using the \texttt{CZM-Hybrid} model, lending support to our hypothesis that the strength surface can significantly influence crack propagation under non-mode I loading. A tensile-to-shear crack transition may occur as a material's compressive strength to tensile strength ratio decreases. However, caution must be exercised in drawing conclusions from this numerical study since a particular form of strength surface has been assumed. A further combined experimental and numerical investigation is required to understand this transition fully.

\subsection{Mixed-mode benchmark of Yosibash and Mittleman \cite{yosibash2016}}

Yosibash and Mittleman \cite{yosibash2016} developed a full three-dimensional benchmark problem to validate phase-field models. Their experimental setup employed four-point bending tests on V-notched specimens with varying notch inclination angles. This configuration generated combined Mode I+II+III loading conditions, resulting in continuously curved crack propagation paths. The researchers systematically measured the crack initiation angles relative to two reference planes as cracks emerged from the V-notches. In our study, we implement the three phase-field models to simulate this benchmark problem, focusing specifically on two of the reported notch inclination angles. We then compare our predicted crack initiation angles with the experimental measurements documented in the benchmark study.

\begin{figure}[h!]
	\centering
	\includegraphics[width=5.5in]{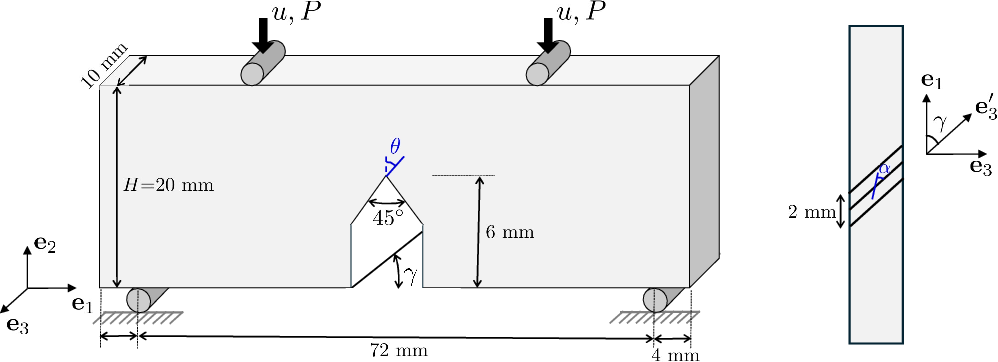}
	\caption{Schematic of the four-point bending experiments with V- notches inclined at an angle of $\gamma$ with the horizontal axis ${\bf e}_1$. The nucleated crack forms an angle $\theta$ with the vertical axis ${\bf e}_2$ as shown in the left figure, and it forms an angle $\alpha$ with an axis parallel to the notch $\textbf{e}^{\prime}_3$ as seen in the figure on the right which displays a view of the specimen from the bottom.}\label{Fig8}
\end{figure}

Figure \ref{Fig8} presents a schematic representation of the four-point bending configuration. The V-notch is oriented at angles of $\gamma=45^{\circ}$ or $60^{\circ}$ relative to the horizontal axis $\textbf{e}_1$. During loading, a crack initiates at an angle $\theta$ with respect to the $\textbf{e}_2$ axis in the $\textbf{e}_1$-$\textbf{e}_2$ plane, while forming an angle $\alpha$ with the $\textbf{e}^{\prime}_3$ axis in the $\textbf{e}_1$-$\textbf{e}_3$ plane. These angles are measured numerically after the crack has propagated by a small distance in the $\textbf{e}_2$ direction. To evaluate potential regularization length effects on the initiation angles, they are calculated at crack lengths of 0.5 mm and 1.0 mm. No significant dependence on regularization length is observed.
The graphite material investigated experimentally by Hug et al. \cite{yosibash2020} has the following mechanical properties: Young's modulus $E=12.44$ GPa, Poisson's ratio $\nu=0.2$, fracture toughness $G_c=118$ N/m, and tensile strength $\sts=48$ MPa. The compressive strength was not reported in the experimental work. Recent studies by Hu et al. \cite{dolbowyosibash2022} have demonstrated that the strength surface can significantly influence the critical nucleation stress in such configurations. For the present analysis, we adopt a compressive strength value of $\scs=150$ MPa---a detailed investigation of strength surface effects is beyond the scope of this work. It should be noted that there exists a discrepancy in the reported elastic modulus between two relevant studies \cite{yosibash2016, yosibash2020}; we utilize the value from the more recent work in our analysis.

\begin{figure}[h!]
	\centering
	\includegraphics[width=6.5in]{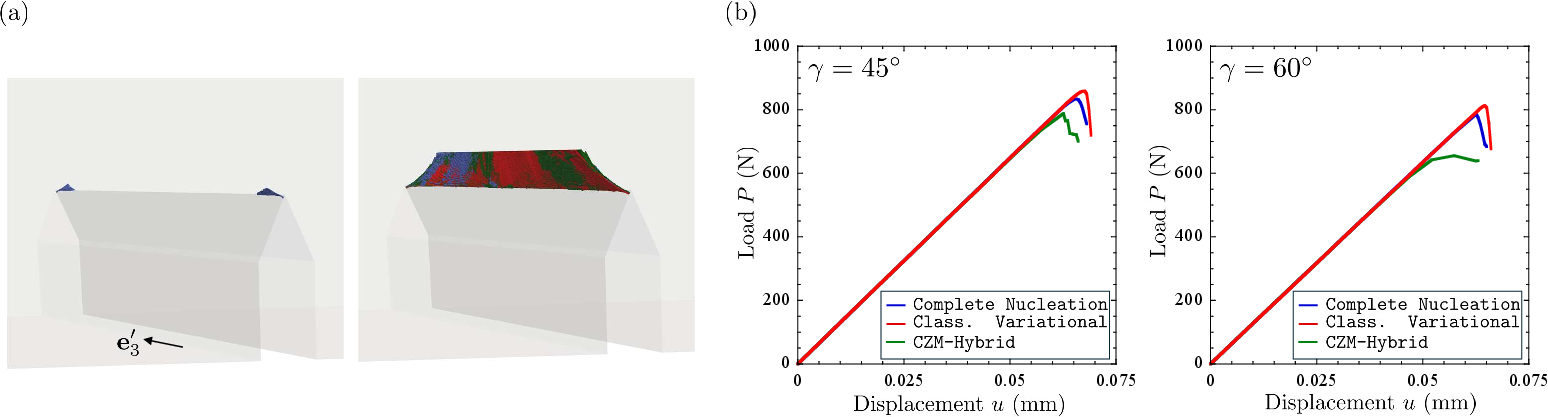}
	\caption{(a) Crack path predicted by the phase-field models for the mixed-mode four-point bending test. (b) Plot of the predicted load $P$ as a function of the applied displacement $u$ by the three phase field models for notch inclination angle $\gamma=$45$^{\circ}$ and 60$^{\circ}$. }\label{Fig9}
\end{figure}
\begin{figure}[h!]
	\centering
	\includegraphics[width=4.5in]{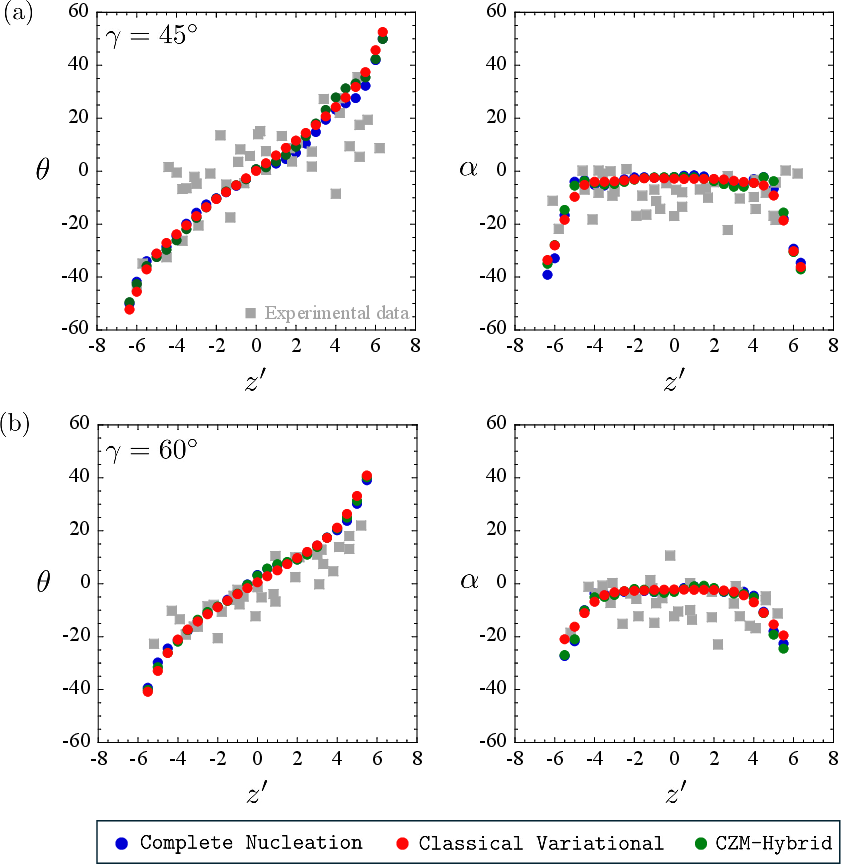}
	\caption{Crack inclination angles $\theta$ and $\alpha$ after a 0.5 mm extension from the V-notch in the $\textbf{e}_2$ direction as predicted by the three phase-field models and compared with experimental measurements for the notch inclination angle (a) $\gamma=$45$^{\circ}$ and (b) $\gamma=$60$^{\circ}$. The angles are plotted with respect to the depth of the notch, $z'$, in the direction of the initial inclination of the V-notch, denoted $\textbf{e}^{\prime}_3$ in Fig.~\ref{Fig8}.}\label{Fig10}
\end{figure}
\begin{figure}[h!]
	\centering
	\includegraphics[width=4.5in]{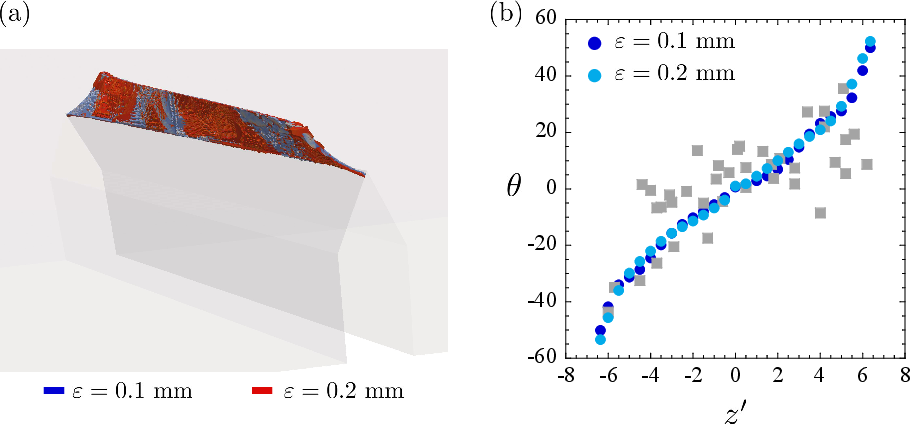}
	\caption{Comparison of the predictions for the four-point bending test from the \texttt{Complete Nucleation} phase-field model for two values of regularization length $\varepsilon=0.1$ mm and 0.2 mm. (a) Crack path and (b) inclination angle $\theta$ as a function of the depth of the notch, $z'$, in the direction of inclination, $\textbf{e}^{\prime}_3$.}\label{Fig11}
\end{figure}

The phase field simulation results are visualized through contour plots with the three models in Fig.~\ref{Fig9}(a), which demonstrates the crack propagation pattern. Initial crack nucleation occurs at the specimen edges, subsequently progressing to nucleate in the central region of the sample. Fig.~\ref{Fig9}(b) presents a comparative analysis of the load-displacement responses obtained from the three phase-field models. The \texttt{Complete Nucleation} and \texttt{Classical Variational} approaches yield nearly identical load-displacement responses for both inclination angles examined. In contrast, the \texttt{CZM-Hybrid} model consistently predicts lower peak forces.
Fig.~\ref{Fig10} illustrates the computed inclination angles $\theta$ and $\alpha$ from all three models, with experimental validation data included for comparison. The numerical predictions from each model demonstrate strong agreement with one another and correlate well with the experimental measurements. The simulations are performed using the regularization lengths $\varepsilon=0.1$ mm for both the \texttt{Complete Nucleation} and \texttt{CZM-Hybrid} models, and $\varepsilon=l_{\rm ch} \approx 0.24$ mm for the \texttt{Classical Variational} model.
To validate the length scale independence of the \texttt{Complete Nucleation} model, additional calculations are conducted using $\varepsilon=0.2$ mm. Fig.~\ref{Fig11}(a) depicts the superimposed crack isovolume ($v<0.1$) obtained from simulations using both regularization lengths, while Fig.~\ref{Fig11}(b) confirms the consistency of the crack inclination angle $\theta$ across both cases.

\subsection{Notched plates under uniaxial tension, biaxial tension and pure shear}

The analyses presented in the previous subsections demonstrate that the \texttt{Complete Nucleation} model effectively characterizes crack propagation under both mode II and mode III shear loading conditions. To further validate the model's capacity to exhibit Griffith-type fracture behavior under generalized loading conditions, we conduct simulations of notched plates subject to three distinct loading scenarios: uniaxial tension, biaxial tension, and pure shear loading. For configurations where the plate dimensions substantially exceed the notch size, and the notch dimensions are significantly larger than the intrinsic fracture length scales, analytical solutions for critical stress are available in the literature for comparisons with predictions from the three phase-field models. In cases where the notch dimensions are comparable to the intrinsic length scales, we perform comparative analyses between the three models. We again employ the graphite material properties specified in Section 3.1. The parameter $\varepsilon$ is assigned a value of 0.1 mm for both the \texttt{Complete Nucleation} and \texttt{CZM-Hybrid} models, while for the \texttt{Classical Variational} model, we set $\varepsilon=l_{\rm ch} \approx 0.46$ mm. Undamaged notch boundary conditions are used.

\begin{figure}[h!]
	\centering
	\includegraphics[width=5in]{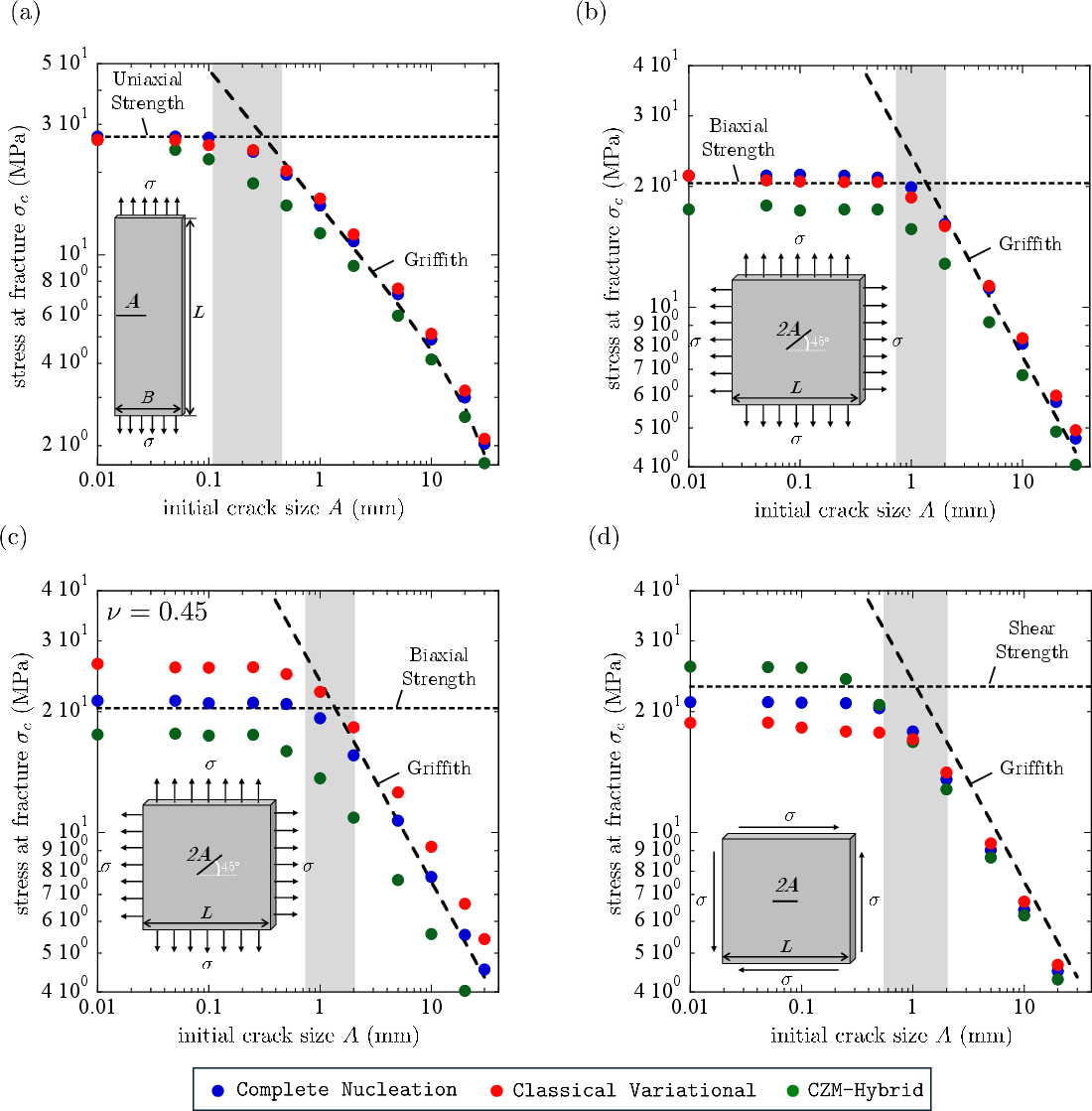}
	\caption{Predictions from the three phase-field models for notched plates subjected to multiaxial loading illustrating the transition from Griffith-dominated to strength-dominated nucleation of fracture as the size size $A$ of the crack decreases from the large to small. The results show the critical stress $\sigma_c$ at which fracture nucleates compared with analytical predictions of nucleation based on Griffith theory of fracture (dashed line) and based on strength (densely dashed horizontal line). The intermediate region surrounding the intercept of the two lines is shown in gray. The results are included for plates made out of graphite subjected to (a) uniaxial tension, (b) and (c) biaxial tension, and (d) pure shear. In part (c), the Poisson's ratio $\nu$ of graphite is altered to 0.45, keeping all other values the same. }\label{Fig12}
\end{figure}

The critical stress results for a single-edged notch plate under tensile loading are presented in Fig.~\ref{Fig12}(a), which compares the three models across notch lengths ranging from 0.01 to 30 mm. The \texttt{Complete Nucleation} and \texttt{Classical Variational} models demonstrate strong agreement throughout the entire range of notch lengths investigated. Moreover, both models exhibit good agreement with analytical predictions for Griffith crack propagation when notch lengths exceed 1 mm ($A>1$ mm). The agreement would be even better if damaged notch boundary conditions are used. As the notch length approaches zero, the critical stress values from both models converge asymptotically to the uniaxial tensile strength of the material. In contrast, the \texttt{CZM-Hybrid} model consistently yields lower predictions for the critical stress compared to the other two approaches.

The critical stress predictions for a center-notched plate under biaxial tension, obtained using the three models, are presented in Fig.~\ref{Fig12}(b). Both the \texttt{Complete Nucleation} and \texttt{Classical Variational} models exhibit good agreement across the entire range of notch lengths. However, it is important to note that the \texttt{Classical Variational} approach cannot independently capture the biaxial strength of the material. For graphite, which has a low Poisson's ratio $\nu=0.13$, the biaxial strength predicted by the \texttt{Classical Variational} model, as per the equation (\ref{strength-surface-star-convex}), closely aligns with the experimental value.
To further investigate, calculations are repeated for a material with a higher Poisson's ratio $\nu=0.45$, while keeping all other material properties identical. Assuming the experimental biaxial strength remains unchanged, the \texttt{Classical Variational} model is found to significantly overpredict the biaxial response for small notch lengths $A<2$ mm as illustrated in Fig.~\ref{Fig12}(c); see \cite{kamarei2025nucleation} for more discussion on this aspect.  By contrast, the \texttt{Complete Nucleation} model exhibit a response independent of $\nu$,  accurately capturing the biaxial behavior across the entire range.
The \texttt{CZM-Hybrid} model, on the other hand, again underpredicts the critical nucleation stress. Moreover, the accuracy of its predictions further deteriorated for the higher Poisson's ratio $\nu=0.45$ as clearly observed from Fig.~\ref{Fig12}(b) and Fig.~\ref{Fig12}(c).

The results for a center-notched plate under pure shear loading are presented in Fig.~\ref{Fig12}(d). In this case, the critical stress predictions from the \texttt{Complete Nucleation} and \texttt{Classical Variational} models deviate from the analytical solution, likely due to the assumption of mode II crack propagation in the analytical calculation. Nonetheless, the predictions from both models are close for $A>1$ mm. For $A<1$ mm, the behavior differs significantly. The \texttt{Classical Variational} model severely underestimates the shear strength. The \texttt{Complete Nucleation} model also does not fully converge to the exact shear strength, though it is expected to do so with a further reduction in the value of $\varepsilon$.

\subsection{Double-edge notched plate under mixed mode loading}

\begin{figure}[h!]
	\centering
	\includegraphics[width=6.5in]{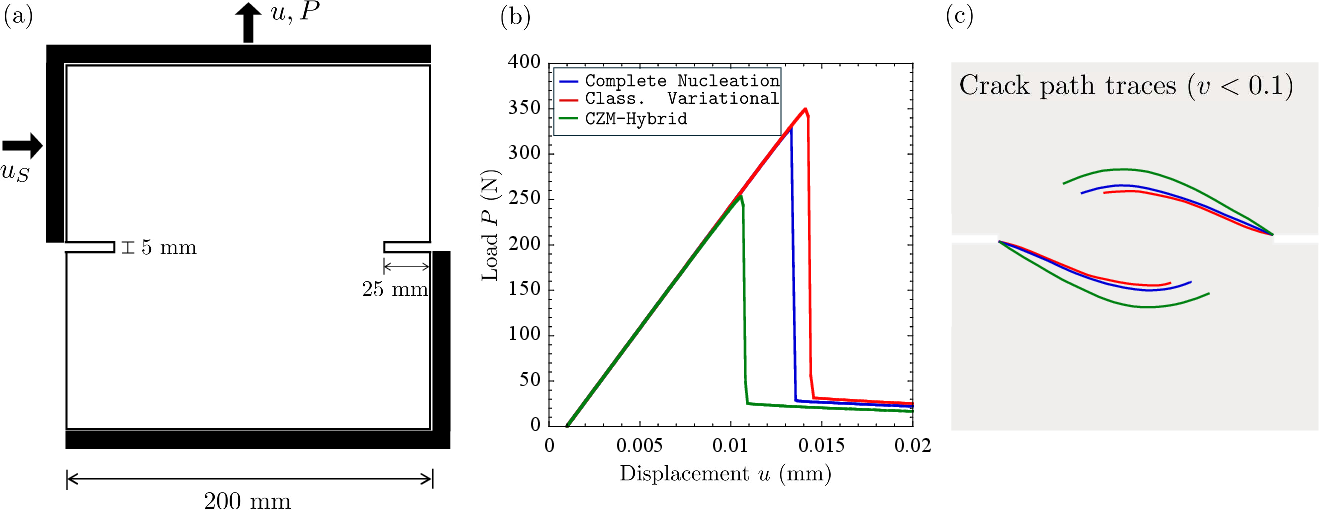}
	\caption{(a) Schematic of the double-edge notched plate under combined shear and tensile loading. (b) Comparison of the load-displacement $P$-$u$ plots for the three phase field models. (c) Comparison of the predicted crack paths.}\label{Fig13}
\end{figure}

We next simulate an experiment designed to study mixed-mode I+II fracture, also known as the Nooru-Mohamed test \cite{nooru1992dent, nooru1993experimental}. In this setup, a double-edged notched plate is affixed to a rigid steel frame, as illustrated in Fig.~\ref{Fig13}. The steel frame is initially deformed horizontally, followed by vertical deformation resulting in a combined tension+shear loading. Similar experimental configurations have been widely reported in the literature including studies conducted under biaxial tension+shear, particularly for concrete and other cement based material; see \cite{carpiuc2017DENT} and the references therein. In this work, we utilize a brittle material behavior and evaluate the predictions of the three phase-field models under consideration. It has been simulated with the \texttt{CZM-Hybrid} model in \cite{wu2020vonMises}.

The following mechanical properties are adopted from the literature \cite{dumstorff2007concrete} for concrete: Young's modulus $E=30$ GPa, Poisson's ratio $\nu=0.2$, and tensile strength $\sts=5$ MPa. A more brittle concrete is employed in the simulations, characterized by a fracture toughness, $G_c$, of 11 N/m. The resulting characteristic fracture length is $l_{\rm ch}\approx$ 5 mm and the fracture behavior can be still approximately regarded as brittle instead of quasi-brittle. We adopt a regularization length of $\varepsilon=5$ mm with all three models. While the experiments are typically force-controlled, we run a displacement controlled simulation in which the shear displacement $u_S$ is first applied and then kept fixed, as the vertical displacement $u$ is incrementally increased. 

The load-displacement, $P-u$, responses for the three models are presented in Fig.\ref{Fig13}(b), while the corresponding crack paths are shown in Fig.\ref{Fig13}(c). The \texttt{Complete Nucleation} and \texttt{Classical Variational} models demonstrate reasonable agreement in both the load-displacement behavior and the crack path prediction. In contrast, the \texttt{CZM-Hybrid} model underpredicts the peak force associated with crack nucleation. This discrepancy also influences the crack path, which exhibits a greater tilt compared to the paths predicted by the other two models.

\subsection{Plate with inclined notch under compression}\label{inclinednotch}

Finally, we examine the extensively studied problem of a plate with an inclined notch under compression using the three models. Experimental observations show the nucleation of a wing-shaped tensile crack emanating from the pre-existing crack, followed by the formation of secondary tensile and shear cracks. Recent studies on a similar configuration involving two inclined notches have demonstrated that the \texttt{Complete Nucleation} approach can predict both wing-crack nucleation and the subsequent formation of shear cracks between the notches.
That work also showed that the \texttt{Classical Variational} approach, employing volumetric-deviatoric or spectral splits that only account for tensile strength, fails to accurately predict the observed crack growth. In this study, we revisit the inclined notch problem with the \texttt{Classical Variational} model using the star-convex split. Star-convex split does provide an asymmetry in tensile and compressive strength; however, the strength surface representation is inadequate with unusually low strength in shear and adjacent states (Fig.~\ref{Fig1}). We seek to evaluate whether it is only the asymmetry in tensile/compressive strength that matters or the entire surface. Moreover, we investigate whether the \texttt{CZM-Hybrid} model, which does not employ an energy split similar to the \texttt{Complete Nucleation} model, can also predict correctly the crack growth.

\begin{figure}[h!]
	\centering
	\includegraphics[width=6in]{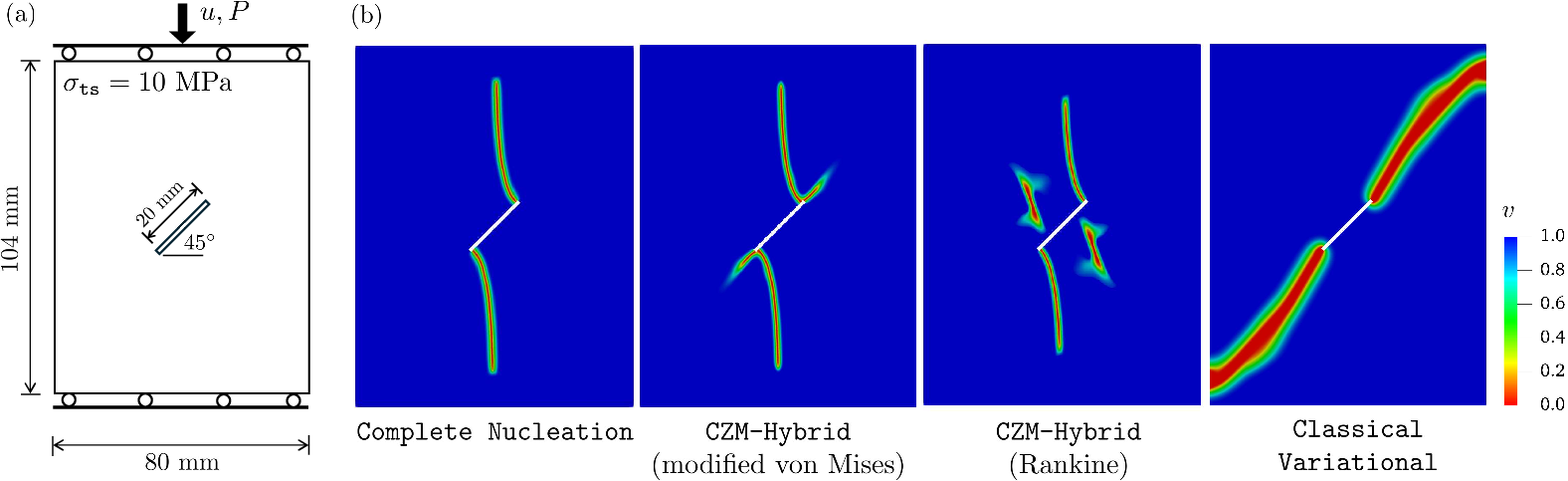}
	\caption{(a) Schematic of the plate with inclined notch under compression. (b) Contour plots of the phase field $v$ as predicted by the \texttt{Complete Nucleation}, \texttt{Classical Variational} and \texttt{CZM-Hybrid} models. The \texttt{Classical Variational} model uses the star-convex split. The \texttt{CZM-Hybrid} predictions are shown with both modified von Mises and Rankine models.}\label{Fig14}
\end{figure}

The material properties of marble are adopted from \cite{jiefan1990marble, wick2022}: $E=63.5$ GPa, $\nu=0.21$, $G_c=11$ N/m and $\scs=175$ MPa. The tensile strength $\sts$ of marble was not specified in the referenced papers; however, it is generally reported to range between 2 and 30 MPa in various sources. We first adopt a value of $\sts=10$ MPa. The regularization length is chosen to be $\varepsilon=1$ mm for both the \texttt{Complete Nucleation} and \texttt{CZM-Hybrid} models, and $\varepsilon=l_{\rm ch} \approx 2.6$ mm for the \texttt{Classical Variational} model.
The simulation results are presented in Fig.~\ref{Fig14}. The \texttt{Classical Variational} model with the star-convex split predicts an incorrect crack path. However, unlike the volumetric-deviatoric split, it does not lead to the formation of a compressive crack. Rather, it initiates two shear cracks, similar to observations in Section \ref{Mode2&3}.  In contrast, both the \texttt{Complete Nucleation} and \texttt{CZM-Hybrid} models predict the nucleation of wing cracks along identical paths. The cracks propagate stably towards the top and bottom boundaries, aligned with the direction of the applied compression.
For this value of tensile strength, the \texttt{Complete Nucleation} model does not predict the nucleation of any secondary cracks. However, the \texttt{CZM-Hybrid} model does predict additional crack nucleation. When using the modified von Mises criterion, a shear crack forms, causing the simulation to rapidly lose stability. Conversely, when the Rankine equivalent stress criterion is applied, a secondary backward tensile crack is observed to nucleate.
These findings underscore the sensitivity of crack growth behavior, particularly for secondary cracks, to subtle differences in the material strength surface.

\begin{figure}[h!]
	\centering
	\includegraphics[width=6in]{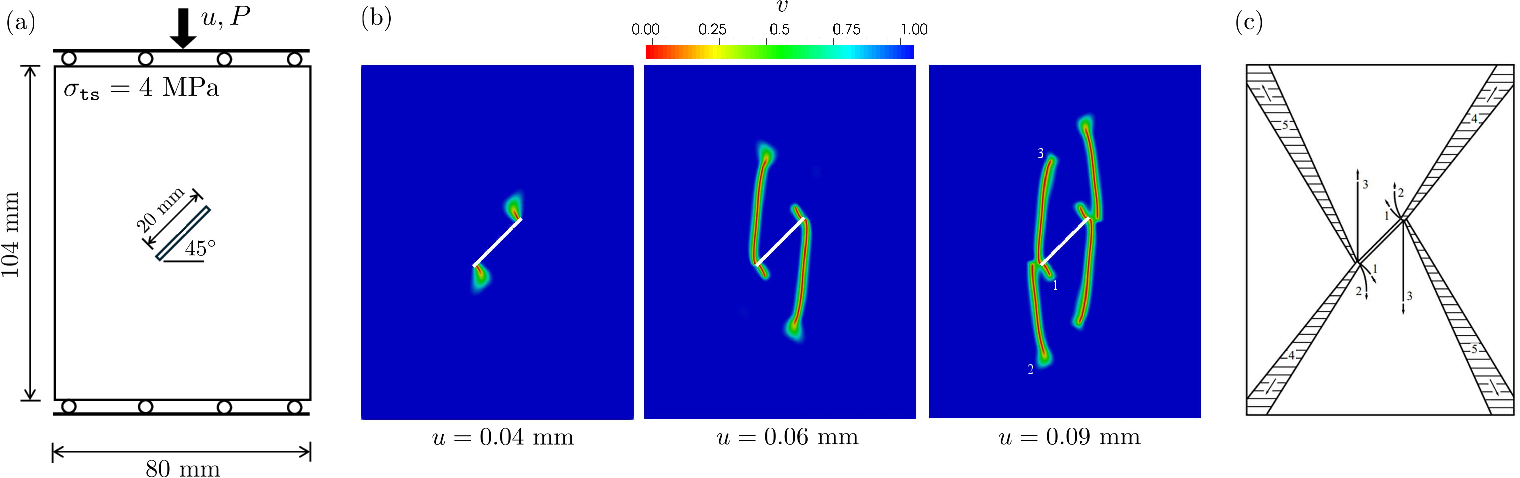}
	\caption{Predictions of the \texttt{Complete Nucleation} model for the case when the uniaxial tensile strength of the material is low, $\sigma_{\texttt{ts}}=4$ MPa. (b) Contour plots of the phase field $v$ at different displacements. (c) Experimental result for a marble plate \cite{jiefan1990marble}. The failure traces are labeled 1, 2 and 3 representing the primary forward tensile cracks, secondary forward tensile cracks and backward tensile cracks respectively, while the shaded areas labeled 4 and 5 represent the forward shear belts and backward shear belts, respectively.}\label{Fig15}
\end{figure}

We also simulate the problem using the \texttt{Complete Nucleation} model with a lower tensile strength $\sts=4$ MPa to gain better consistency with experimental results shown in Fig.~\ref{Fig15}(c). Under these conditions, the model predicts the nucleation of additional cracks. Specifically, both a backward tensile crack and a secondary forward tensile crack are observed, as shown in Fig.~\ref{Fig15}(b), which is consistent with experimental observations in marble. However, experiments also show the formation of shear bands which are not seen in the simulations. This suggests that there is some quasi-brittle fracture physics that will need to be added to the model.  A different strength surface may also be needed to more accurately approximate the strength envelope for marble.

\section{Summary and final comments}\label{Sec: Final Comments}

The six sets of comparisons presented in this study offer a comprehensive evaluation of the three classes of phase-field models for brittle fracture that incorporate strength surfaces, specifically focusing on crack growth under mode II, mode III, and mixed-mode loading conditions. These three classes of models incorporate strength into the variational framework for fracture propagation using distinct approaches.

The \texttt{Complete Nucleation} model incorporates the strength surface directly as a stress-based driving force, albeit at the expense of variational consistency. Comparative analyses demonstrate that this model effectively predicts crack growth under mode II, mode III, and mixed-mode loading across a wide range of material behaviors and geometries in accordance with Griffith's theory of crack propagation.
When combined with findings from previous studies on crack nucleation, these results provide compelling evidence that the \texttt{Complete Nucleation} model is capable of accurately predicting both the initiation and propagation of fractures in elastic brittle materials. Note, however, that while the global loading studied in this work is non-mode I, the crack locally tends to propagate in a mode I fashion. Most experimental setups inadvertently introduce mixed-mode loading conditions, so it is difficult to validate the model for cases where cracks propagate in pure mode II or mode III. Another challenge is that the material may not behave in a brittle manner under these conditions. However, this needs to be further studied.  Furthermore, a mathematical proof establishing the convergence of the crack growth behavior emerging from this model to Griffith's theory in the limit as $\varepsilon \searrow 0$ remains desirable. Such a proof would also include the mathematical formalization of the idea of tension-compression asymmetry in crack propagation arising from physically consistent strength surfaces that is now well-supported by numerical results.

The \texttt{Classical Variational} model incorporates material strength indirectly through an energy split. It is well-established that this approach fails to accurately capture the strength surface and, consequently, the nucleation behavior within the bulk. The comparisons presented in this study further highlight limitations in predicting crack nucleation from a large crack. Specifically, the results demonstrate that a star-convex energy split, which has been argued to be superior to other splitting methods, unexpectedly leads to the nucleation of a compressive crack under simple shear loading.
Moreover, the star-convex split exhibits a tendency for shear crack nucleation in various scenarios, including mode II loading with high Poisson's ratios, mode III loading, and inclined notch compression problems. In these cases, shear crack nucleation precedes tensile crack formation with the variational model, unlike the other two models for realistic strength values. Additional experimental validation is required for some of these cases to understand whether the variational model or the other two models are giving experimentally consistent results. 
Closely connected are the observations in classical fracture mechanics literature that materials may have different critical energy release rates in different modes.
% , or the crack path predicted by the maximum hoop stress criterion and maximum energy release rate criteria may be different. 
Still, these findings emphasize the need to incorporate the full strength surface into the model, rather than merely accounting for the asymmetry between tensile and compressive strengths. They further emphasize the need to treat the different issues within the phase field formulations in a unified manner.
Although individual issues identified in this study could potentially be mitigated by employing alternative energy splits, a unique energy split capable of addressing all these challenges while satisfying other fundamental properties—such as those discussed in \cite{LorenzisMaurini2024nucleation}—appears to be currently unavailable. A reassessment of the whole approach will prove useful.
Nevertheless, the \texttt{Complete Nucleation} and \texttt{Classical Variational} models yield nearly identical predictions for mode I+II and mode I+III problems dominated by tension.

The \texttt{Cohesive Zone Hybrid} model incorporates the strength surface by modifying the energy function in the equation governing the evolution of phase field. It employs an asymmetric energy function for tension and compression, designed based on the form of the strength surface. While this approach cannot perfectly capture the strength surface—particularly under hydrostatic tension or compression—it effectively approximates the general shape of the strength surface for many materials. This is a notable improvement over the \texttt{Classical Variational} model. As a result, the \texttt{Cohesive Zone Hybrid} model does not exhibit the issues associated with the \texttt{Classical Variational} approach and generally produces results that are qualitatively consistent with the \texttt{Complete Nucleation} model across six comparative studies.
However, the \texttt{Cohesive Zone Hybrid} model suffers from a significant quantitative limitation: it predicts an incorrect effective fracture toughness. This discrepancy arises because the introduction of the asymmetric energy function, intended to approximate the strength surface, disrupts the Griffith energy competition. It remains unclear whether a correction---similar to the one employed in the \texttt{Complete Nucleation} model---could be applied to address this issue.

\section*{Acknowledgements}

\noindent The authors would like to acknowledge the financial support from the start-up fund provided by the Georgia Institute of Technology and from the National Science Foundation, United States, through the grant CMMI-2404808. The computations reported here were conducted through research cyberinfrastructure resources and services provided by the Partnership for an Advanced Computing Environment (PACE) at the Georgia Institute of Technology. This work began following spirited discussions at the 2024 World Congress on Computational Mechanics, for which we thank the conference organizers. 

\bibliographystyle{elsarticle-num-names}
\bibliography{ref}

\end{document}